\documentclass[]{emulateapj}
\usepackage{natbib}
\usepackage{graphicx}

\newcommand{\init}{$p8$}
\newcommand{\rate}{$r$}
\newcommand{\cutoff}{$c$}
\newcommand{\fraction}{$f$}
\usepackage{mathrsfs}
\usepackage{comment}
\usepackage{gensymb} 

\usepackage[T1]{fontenc}
\usepackage{ae,aecompl}

\usepackage{amsmath}
\usepackage{amsfonts}
\usepackage{amssymb}
\usepackage{comment}
\usepackage[dvipsnames]{xcolor}
\usepackage[figuresleft]{rotating}

\usepackage[%
        ]{hyperref}

\begin{document}

\hypersetup{}


\title{Ages of ``singles" versus ``multis": Predictions for dynamical sculpting over Gyr in the \textit{Kepler} Sample}

\author{Christopher~Lam\altaffilmark{1} and Sarah~Ballard\altaffilmark{1}}

\altaffiltext{1}{University of Florida Department of Astronomy, 211 Bryant Space Science Center, Gainesville, FL 32611, USA; c.lam@ufl.edu}


\begin{abstract}
The sample of host stars with multiple transiting planets has illuminated the orbital architectures of exoplanetary systems. These architectures may be shaped mostly by formation conditions, be continually sculpted by ongoing dynamical processes, or both. As more studies place planet occurrence within a galactic context, evidence has emerged for variable planet multiplicity over time. In this manuscript, we investigate the use of transit multiplicity as a tool to constrain longer-timescale ($>$1 Gyr) dynamical sculpting. First, with a suite of injection-and-recovery tests, we quantify sensitivity to sculpting laws across different regimes. We employ a forward modeling framework in which we generate synthetic planetary systems, according to a prescribed sculpting speed and timescale, around the FGK dwarfs studied by the \textit{Kepler} Mission. Some sculpting scenarios are hypothetically detectable in the \textit{Kepler} sample, while others can be disfavored from \textit{Kepler} transit statistics alone. Secondly, we apply our analysis to reverse-engineer the sculpting laws consistent with the true yield from \textit{Kepler}. We confirm the present-day fraction of host stars containing dynamically cool ``systems with tightly-packed inner planets" (STIPs) is 4-13\%. A variety of Gyr-timescale sculpting laws successfully predict the transit multiplicity of the Kepler sample, but none of these laws succeeds in also producing a detectable trend with transit multiplicity and stellar age. An improvement to measured stellar age precision may help uncover such a sculpting law, but nevertheless reflects limitations in transit multiplicity as an observable. Therefore other phenomena, apart from Gyr-timescale dynamical sculpting, may be required to explain the \textit{Kepler} yield.
\end{abstract}

\section{Introduction}
Detected planet yields from transit surveys encode a wide range of demographic information. While each of the thousands of planetary systems discovered by NASA's {\it Kepler} mission \citep{borucki_kepler_2010, borucki_characteristics_2011, borucki_characteristics_2011-1, batalha_planetary_2013} contribute to our understanding of raw planet occurrence, systems with more than one transiting planet (also known as ``multi-transiting systems", or ``multis") enable investigations of system architecture (e.g. \citealt{fabrycky_architecture_2014}; see \citealt{Dawsontextbook} for a full review). 

The subfield of multi-transit systems enabled by \textit{Kepler} motivated the introduction of terminology to quantify ensemble architectural properties. “Dynamical temperature” \citep{tremaine_statistical_2015}, for example, describes the ensemble dynamical state of a planetary system using a statistical mechanics framework. Orbital eccentricity $e$, orbital mutual inclination between planets $i$, and spacing between planets are all low in systems that are ``dynamically cool.” These systems also tend to host more close-in planets on average \citep{lissauer_architecture_2011, fang_architecture_2012, limbach_exoplanet_2015, millholland_evidence_2021}. So-called ``systems of tightly-packed inner planets" (STIPs, per \citealt{Ford14}) may or may not coexist with further-out companions \citep{becker_effects_2017, zawadzki_migration_2022}. Planets in ``dynamically hot” systems, in contrast, reside at wider spacings from one another and with higher eccentricities and mutual inclinations. The ``angular momentum deficit" (AMD, per \citealt{Laskar17}) is also a useful ensemble statistic, quantifying the difference between the actual angular momentum in a system of planets, and the idealized maximum angular momentum if all planets possessed eccentricity $e=0$. In this sense, systems with high angular momentum deficit are also ``dynamically hot." 

The transit multiplicity distribution (that is, the number of stars hosting $n$ transiting planets) allows us to reverse-engineer typical dynamical temperatures of planetary systems in the \textit{Kepler} yield. In a sample with more dynamically cool systems, all else being equal, the transit multiplicity distribution will contain more multi-transiting systems \citep{fabrycky_architecture_2014}. The term ``{\it Kepler} Dichotomy" \citep{lissauer_architecture_2011} emerged as an early heuristic descriptor for the seeming over-abundance of observed single-transiter systems, when compared to the number of multi-transiter planetary systems. Some studies have gone on to employ a mixture model to explain the phenomenon, typically invoking one type of planetary system to furnish the ``multis", and another to explain the excess ``singles" \citep{lissauer_architecture_2011, fang_architecture_2012, johansen_can_2012, hansen_testing_2013, ballard_thekeplerdichotomy_2016, dawson_correlations_2016}. Other studies determined that the effect can be well-modeled by a single distribution given the right set of assumptions \citep{zhu_about_2018, he_architectures_2019, millholland_evidence_2021, zawadzki_migration_2022} (it may also be at least partly spurious, per \citealt{zink_accounting_2019}). Importantly, the statistics of multi-transiting planets from the \textit{Kepler} mission bear many theoretical predictions for how the observed transit multiplicity distribution varies with other quantities. Properties such as $e$, $i$, mutual Hill spacing, frequency of transit timing variations, and even atmospheric properties vary together with transit multiplicity, though the hypothesized provenance of the distribution varies \citep{johansen_can_2012, hansen_testing_2013,   fabrycky_architecture_2014, xie_exoplanet_2016, eylen_orbital_2019, ballard_thekeplerdichotomy_2016, dawson_correlations_2016, he_architectures_2019, he_architectures_2020, millholland_tidally-induced_2019}. 

In this manuscript, we explore the extent to which \textit{evolution} of dynamical temperature over $>1$ Gyr timescales is observable in the \textit{Kepler} yield. It is as-yet unclear to what extent that distribution is set at formation, and to what extent it evolves over time. Our study is motivated by the recent findings, enabled by the Gaia Mission \citep{Gaia_2016}, that planet occurrence may vary as a function of the galactic properties of the host star. The changes in planet occurrence with stellar kinematic history \citep{Winter20, dai_planet_2021, chen_planets_2021, Longmore21, bashi_exoplanets_2022} or galactic height \citep{zink_scaling_2023} may (1) be linked to galactic environment directly (e.g. \citealt{Cai17, Ndugu22}), (2) reflect underlying relationships with stellar metallicity or age (in this sense, galactic properties are proxies for the deterministic relationship, see e.g. \citealt{Nielsen23}), or (3) some combination of these. One of these findings is particularly suggestive of long-term changes in the dynamical temperature of planetary systems: \cite{chen_planets_2021} identified a trend in transit multiplicity as a function of stellar age. Combining spectroscopic characterization of \textit{Kepler} stars together with Gaia kinematics, \cite{chen_planets_2021} found that field stars are generally older than planet host stars ($4.28_{-0.35}^{+0.55}$ Gyr, as opposed to $3.57_{-0.27}^{+0.43}$ Gyr). Suggestively, the hosts to 2 and $\ge3$ transit multi-planet systems are younger still, with hosts to $\ge3$ planets typically $2.84_{-0.29}^{+0.34}$ Gyr old.  \cite{zink_scaling_2023} similarly found that a decrease in planet occurrence out of the galactic midplane is stronger than the expectation from decreased metallicity alone. 

A model in which the dynamical temperature of planetary systems evolves over galactic time could hypothetically reproduce the findings of \citep{chen_planets_2021} and \citep{zink_scaling_2023}. Within this framework, planetary systems would evolve within eccentricity-inclination $(e,i)$ parameter space, starting at birth in a dynamically cooler/lower AMD state, and then evolving to a dynamically hotter/higher AMD state, in a way that ought to be reflected in transit multiplicity. Such an evolution must be sufficiently strong to remove planets from the transiting geometry. We specifically explore here the possibility that planetary systems evolve in dynamical temperature over Gyr timescales, to specifically test whether any fiducial dynamical sculpting model can reproduce the trend in transit multiplicity with stellar age. Constraints placed on dynamical timescales can also be useful in their own right: they can serve as constraints placed on the large parameter space explored by dynamical N-body simulations. Moreover, investigating the evolution of orbital architectures across systems around stars similar to the Sun may not only help constrain the timescales along which life on Earth could have evolved, but also have implications for the search for life in planetary systems around other Sun-like stars, assuming that excessive dynamical heat could perturb orbits into inclinations and eccentricities that lead to heating cycles untenable for the development and survival of life \citep{teixeira_constraints_2022}.
 
This manuscript is organized as follows: in Section \ref{sec:data}, we describe the data sample selection (Section \ref{sec:sample_selection}) and make initial inferences on dynamical sculpting based on the sample (Section \ref{sec:multi_single_ratios}). In Section \ref{sec:methods}, we define our suite of dynamical sculpting models (Section \ref{sec:generating_models}), the planetary system simulation machinery (Section \ref{sec:generating_planets}), the transit and completeness calculations (Section \ref{sec:transits_and_completeness}), and our model evaluation techniques (Section \ref{sec:model_evaluation}).  In Section \ref{sec:results}, we present the results from our model evaluation. We perform a battery of injection-recovery tests to assess our forward modeling pipeline's sensitivity to different dynamical sculpting regimes (Section \ref{sec:injection-recovery}). We discuss these findings and draw inferences about what they can and cannot tell us about dynamical sculpting timescales. We comment on the plausibility of Gyr-timescale dynamical sculpting to explain a trend in transit multiplicity with stellar age. In Section \ref{sec:Conclusions}, we summarize our findings and conclude. 

\section{Data Sample}
\label{sec:data}
\subsection{Sample Selection}
\label{sec:sample_selection}

We select our stellar sample from Table 2 of \citet{berger_gaia-kepler_2020-1}, which cross-matches {\it Kepler} stellar properties with {\it Gaia} Data Release 2. Following the procedure in Section 4.1 in \citet{berger_gaia-kepler_2020}, we perform an initial filter of this sample, removing entries with poor goodness-of-fit to the nearest MIST model grid point (\citet{paxton_modules_2011, paxton_modules_2013, paxton_modules_2015}; \citet{choi_mesa_2016}; \citet{dotter_mesa_2016}), uninformative posteriors (terminal age of main sequence, or TAMS, greater than 20 Gyrs), binary stars (re-normalized unit-weight error, or RUWE, greater than 1.2), and those with uninformative age constraints (that is, stars with fractional age errors greater than the median of 0.56). We then winnow the sample further to retain only FGK stars (3700-7500 K) and remove giants, resulting in a final sample of 67380 stars. Filtering by spectral type first and then by fractional age errors also results in a sample of 67380 stars. We henceforth loosely refer to these stars as ``Sun-like stars", since even though we include a large effective temperature range, the other filters described above leave the sample with a majority of G dwarfs, a significant share of F dwarfs, and a small number of K dwarfs.

\begin{figure}
\includegraphics[width=.45\textwidth]{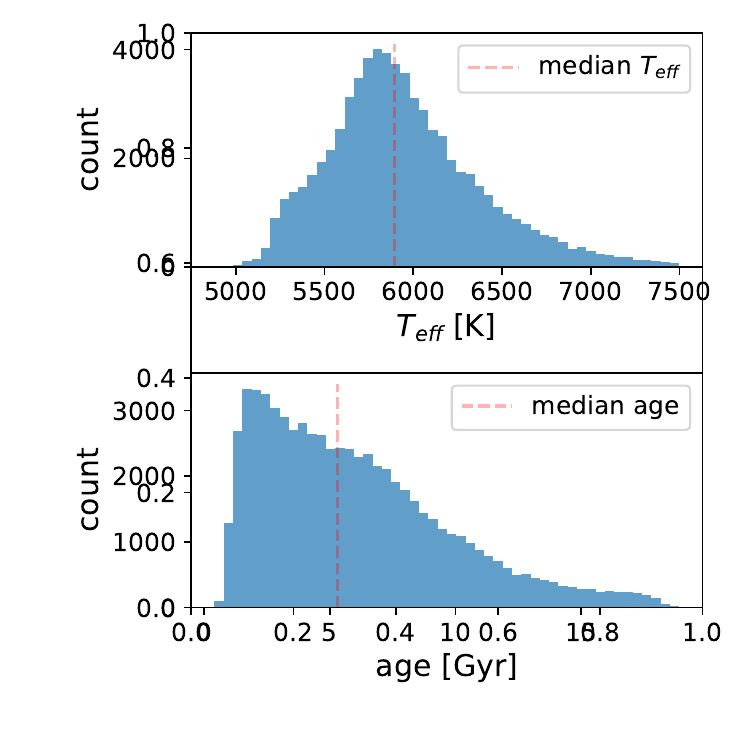}
\caption{Distributions of stellar effective temperature (top) and stellar age (bottom) after all cuts to the sample are made. Vertical, dotted, red lines indicate the medians (5903 K and 5.17 Gyrs, respectively).}
\label{fig:sample_properties}
\end{figure}

Specifically, the cut on age uncertainty biases our sample away from cooler, more slowly-evolving dwarfs. As shown in the stellar effective temperature distribution in Figure \ref{fig:sample_properties}, K dwarfs comprise about 4\% of the sample (compared to $\sim$15\% of the Kepler Input Catalog, per \cite{batalha_selection_2010}).  The bottom, stellar age distribution panel in Figure \ref{fig:sample_properties} shows a sharp drop in systems younger than 1 Gyr, which suggests that forward modeling with this sample should be more robust for ages -- and, therefore, dynamical timescales that operate -- on the order of 1 Gyr and beyond. The median stellar effective temperature and age, indicated by the vertical dotted red lines, are 5903 K and 5.17 Gyrs, respectively. We note that while we choose to employ ages derived from isochrones in \citet{berger_gaia-kepler_2020}, which used v1.2 of the MIST models (\citet{paxton_modules_2011, paxton_modules_2013, paxton_modules_2015}; \citet{choi_mesa_2016}; \citet{dotter_mesa_2016}), \citet{berger_gaia-kepler_2020} validated their stellar ages on cluster, asteroseismic, and kinematic ages and found no significant inconsistencies with the isochrone ages. In Section \ref{sec:injection-recovery}, we demonstrate with injection-recovery tests how well our forward modeling pipeline can perform given isochrone age precisions.

To ensure that our results are robust against the choice of planet vetting procedure, we employ three different planetary status indicators from the Exoplanet Archive when we cross-match planet hosts against the culled sample: koi\_disposition, koi\_score, and fpp\_prob. For the koi\_disposition and koi\_score indicators, we use the Cumulative KOI Table\footnote{Accessed on 2021-03-04, returning 9564 rows}; for the fpp\_prob indicator, we use the False Positive Probabilities Table\footnote{Accessed on 2022-08-25, returning 8054 rows}.

For koi\_disposition, we keep planets marked as `CONFIRMED' or `CANDIDATE' and note each star's planet count. According to the Exoplanet Archive, these are planets that pass the following four false positive tests: the transit dip shape must not be ``Not Transit-Like"; the transit dip shape and depth must not be suggestive of a ``Stellar Eclipse" binary; the star must not exhibit a significant ``Centroid Offset" during the transit, which would suggest a background eclipsing source; and the ephemeris match must not indicate ``Contamination". For the koi\_score and fpp\_prob indicators, we keep planets greater than 0.5 and less than 0.5, respectively. The former indicator is calculated based on the Robovetter tool, which scores a threshold crossing event (TCE) depending on the fraction of Monte Carlo iterations that pass all 28 tests listed in Table 3 of \cite{thompson_planetary_2018}. The fpp\_prob indicator is based on the \texttt{vespa} automated validation pipeline by \cite{morton_efficient_2012} and is obtained directly from the column named, ``Probability to be Any of the Considered Astrophysical False Positive Scenarios (determined by vespa calculation)", in the Kepler False Positive Probabilities Table on the Exoplanet Archive. 

\begin{figure}
\includegraphics[width=.45\textwidth]{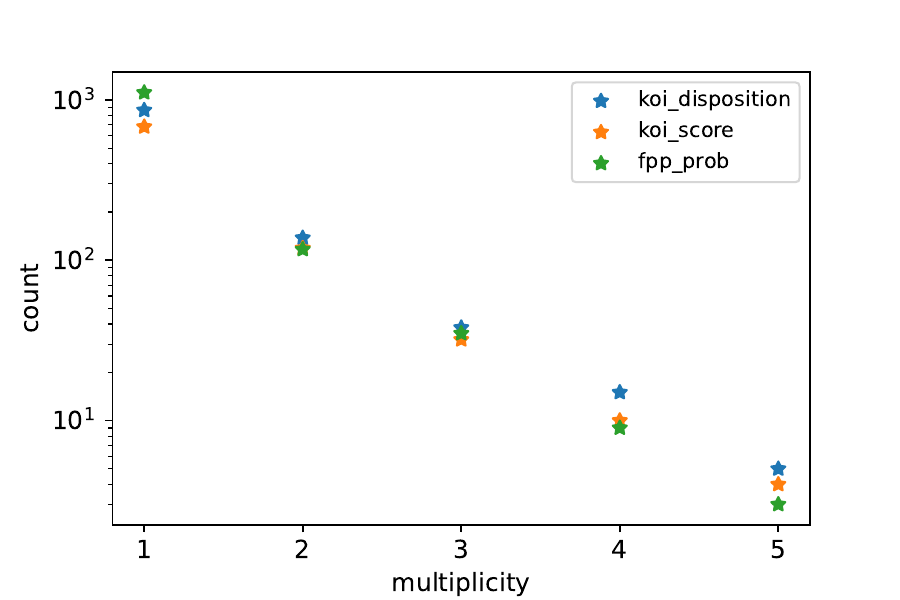}
\caption{Observed {\it Kepler} transit multiplicities using different planet false positive indicators. For ``koi\_disposition", we keep planets marked
as ‘CONFIRMED’ or ‘CANDIDATE’ in the koi\_disposition column from the KOI Cumulative List Table on the Exoplanet Archive. For ``koi\_score", we keep planets with a score of greater than 0.5 in the koi\_score column in the KOI Cumulative List Table on the Exoplanet Archive. Finally, for ``fpp\_prob", we keep planets with a score of less than 0.5 in the ``Probability to be Any of the Considered Astrophysical False Positive Scenarios" column in the Kepler False Positive Probabilities Table on the Exoplanet Archive.}
\label{fig:observed_multiplicities}
\end{figure}

Grouping the data into bins of exoplanet count, for the koi\_disposition indicator, we obtain a distribution of 66355 0-planet systems, 833 1-planet systems, 134 2-planet systems, 38 3-planet systems, 15 4-planet systems, and 5 5-planet systems. For the koi\_score indicator, the observed transit multiplicity is 66588 0-planet systems, 631 1-planet systems, 115 2-planet systems, 32 3-planet systems, 10 4-planet systems, and 4 5-planet systems. For the fpp\_prob indicator, the observed transit multiplicity is 66131 0-planet systems, 1088 1-planet systems, 115 2-planet systems, 34 3-planet systems, 9 4-planet systems, and 3 5-planet systems. These transit multiplicities are shown in Figure \ref{fig:observed_multiplicities}. For this manuscript, we elect to use the koi\_disposition indicator for the ground-truth data against which we compute likelihoods of different dynamical sculpting models. We find no significant differences in our results using either of the other planet disposition indicators (see Section \ref{sec:intact_frac} for further discussion).

\subsection{Changes in observed multi-to-non-multi ratios}
\label{sec:multi_single_ratios}
If movement from low $(e,i)$ space to higher $(e,i)$ occurs over time as shown in \citet{dai_planet_2021}, we would expect fractionally less multi-transiting systems to orbit older stars, compared to younger stars. We assume here that such a change in dynamical temperature would be the dominant driver of an observed discrepancy in compact multis across stellar ages. We perform an initial analysis of the sample, varying the age threshold between so-called ``old" and ``young" systems in 0.5-Gyr increments between 0.5 Gyr and 10 Gyrs. For each threshold, we divide the planet-hosting sample into a ``young" set (stars younger than the threshold) and an ``old" set (stars older than the threshold). We then take the multi-to-non-multi ratios for each subset per age threshold, where a multi is defined as a system having three or more planets within 1 AU, as done in \citet{brewer_compact_2018}. We perform a bootstrap analysis to account for age uncertainty. The stellar age errors in our dataset are asymmetric, however, so in order to draw from these skew distributions, we use Equation 6 from \citet{jontof-hutter_following_2021}, which is adopted from \citet{barlow_asymmetric_2004}. We note that re-sampling stellar ages assuming uniformly distributed ages between the given asymmetric errors makes little difference from assuming normally distributed age errors (for more discussion on the sample's ages, see Section \ref{sec:ages}). 

For each bootstrapped sample, we record the transit multiplicity distribution for ``young" and ``old" stars, noting specifically the ratio of the number of stars with three or more transiting planets ($N_{\textrm{multi}}$) to the number of stars with one or two transiting planets ($N_{\textrm{non-multi}}$). As shown in the top left panel of Figure \ref{fig:multi_ratios_metallicity}, we find that for all age thresholds, there is no significant difference between the observed multi-to-non-multi ratios for ``young" and ``old" systems, at the 2-$\sigma$ level. The scatter at very young thresholds is due to the small sample sizes at this sparsely populated end of the overall sample's age distribution.

\subsection{Metallicity and spectral type considerations}
\label{sec:metallicity}

The chemical evolution of the Milky Way toward increasingly metal-rich stars (e.g. \citealt{Tinsley80}, see \citealt{Matteucci21} for review) means that older Milky Way stars are, on average, more metal-enriched than their younger counterparts. We investigate the extent to which metallicity complicates the comparison between the planet populations of ``young" and ``old" host stars in our sample by first classifying metal-poor stars as those with [Fe/H] less than -0.25, metal-rich stars as those with [Fe/H] greater than 0.25, and Solar-metallicity stars as those with [Fe/H] between -0.25 and 0.25. The \citet{berger_gaia-kepler_2020} cross-match contains both metallicities from \texttt{Gaia} and spectroscopic metallicities from the California Kepler Survey (CKS; \citet{petigura_california-kepler_2017}) for planet hosts. Since the spectroscopic metallicities from CKS are complete and more accurate for the planet hosts in our sample, we use these metallicities to inform the multi-to-non-multi ratio analysis that follows. 

\begin{figure}
\includegraphics[width=.45\textwidth]{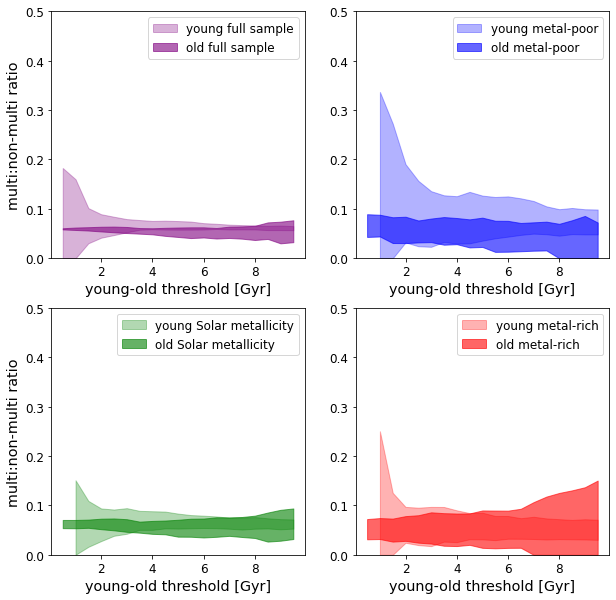}
\caption{We split the {\it Kepler} planet-host sample into young and old populations along different thresholds (x-axis). The y-axis shows the multi-to-non-multi ratio of each ``young" or ``old" population, where multis are systems with three or more planets and non-multis are systems with one or two planets. If there is evidence for sculpting, we would expect the older population to have a lower ratio after a certain young-old threshold. We construct 95th percentile confidence envelopes by bootstrapping 100 times. We observe no significant difference in the multi-to-non-multi ratio between young and old systems, and we find that this lack of age dependence does not change when we disaggregate by stellar metallicity.} 
\label{fig:multi_ratios_metallicity}
\end{figure}

As shown in Figure \ref{fig:multi_ratios_metallicity}, for no metallicity subdivision do we find any significant difference in multi-to-non-multi ratios across the young-old thresholds we test. We repeated the analysis with -0.5 and 0.5 as the metal-poor and metal-rich thresholds, respectively, and also found no significant difference in the multi-to-non-multi ratio for any metallicity bin.

We therefore draw two conclusions from our comparison of the metallicities of ``young" versus ``old" stars. We find (1) no statistical offset in metallicity as a function of the host star age but (2) a modest enhancement in the number of multi-planet systems among the most metal-poor hosts. These conclusions are consistent with previous findings. While \citet{chen_planets_2021} did observe an offset in [Fe/H] of 0.3 between the thin- and thick-disk stars included in \textit{Kepler} sample, the planet sample is drawn almost entirely from the thin-disk (comprising 5\% of the planet hosts considered). They observed no offset in metallicity between field stars and planet hosts as a whole. Indeed, while they found that hosts to $\ge3$ transiting planets are 1.5 Gyr younger on average than field stars, their metallicities were indistinguishable at -0.02$^{+0.17}_{-0.18}$ and -0.02$^{+0.19}_{-0.26}$ respectively \citep{chen_planets_2021}. \citet{zink_scaling_2023} also found that [Fe/H] gradients alone cannot account for the change in planet occurrence with galactic midplane height, and \citet{bashi_exoplanets_2022} found that it cannot account for the change in planet occurrence between thin- and thick-disk populations. Other studies have positively identified a metallicity offset between field stars and the hosts to multi-planet systems, but in the opposite direction predicted by a galactic trend toward multis around younger stars. Rather, \cite{brewer_compact_2018} and \cite{Anderson21} found that compact multi-planet systems are likelier to reside around metal-poor stars. In a galactic sense these ought to be older stars on average, in contrast with the relative youth of multi-planet systems identified by \cite{chen_planets_2021}. 

\begin{figure}
\includegraphics[width=.45\textwidth]{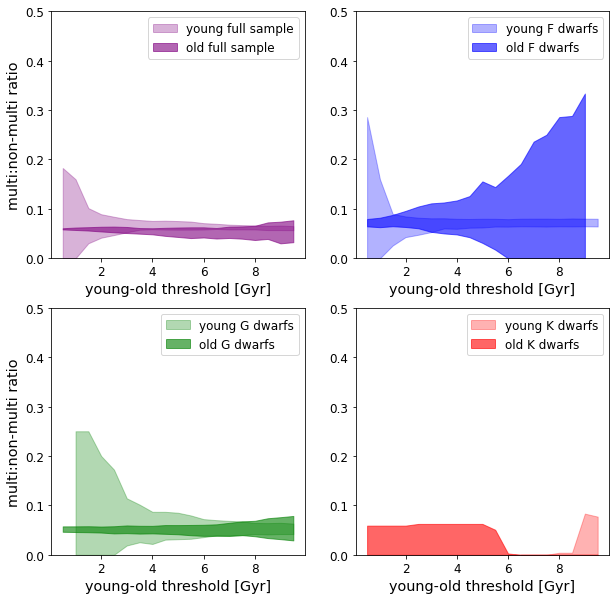}
\caption{We run the same experiment as in Figure \ref{fig:multi_ratios_metallicity}, comparing the multi-to-non-multi ratio of different ``young" and ``old" sub-samples, but now disaggregating by stellar type. We find that disaggregating by stellar type does not change the full-sample picture of a lack of difference in the multi-to-non-multi ratio between young and old systems. Since K dwarfs comprise only a small part of our sample, it is subject to shot noise.} 
\label{fig:multi_ratios_type}
\end{figure}

To account for the role of stellar type in planet multiplicity, we divide our planet host sample into F (hotter than 6000 K), G (5200 - 6000 K), and K (cooler than 5200 K) dwarfs and conduct the same multi:non-multi ratio experiment over each sub-sample. As shown in Figure \ref{fig:multi_ratios_type}, we find that across stellar types F and G there is no significant change in the multi-to-non-multi ratio between ``young" and ``old" stars, for thresholds ranging from 0.5-10 Gyr.

\section{Methods}
\label{sec:methods}

We aim to explore whether a change in planetary system architectures, in the form of increasing dynamical temperature over Gyr timescales, is detectable via the transit multiplicity metric within the \textit{Kepler} sample. Hypothetically, such a phenomenon could reproduce a finding of younger ``multi" systems and older ``singles", separated by Gyr timescales. These characteristics form the basis of a two-class heuristic that we employ here, with originally metastable ``intact" systems becoming  ``disrupted". Within this simplified framework, dynamically cold intact systems will have more planets, more circular orbits, and smaller mutual inclinations. Dynamically hot disrupted systems, in contrast, will have fewer planets, larger eccentricities, and larger mutual inclinations.

\subsection{Generating Dynamical Sculpting Models}
\label{sec:generating_models}

 We first consider the processes that could produce the hypothesized change in transit multiplicity over Gyr timescales. For the purposes of employing transit multiplicity as a metric for dynamical excitation, we require only that this excitation is sufficient to excite planets from a transiting to a non-transiting geometry, which is typically a difference of $\sim$2$^{\circ}$ \citep{fabrycky_architecture_2014-1}. 
 
 Timescales for the dynamical sculpting of planetary systems depend upon the dominant physical mechanism involved, and range from the lifetime of the protoplanetary disk to the lifetime of the star. Because we are specifically testing whether sculpting can provide a Gyr age offset between hosts to single- and multi-transiting systems, we are not considering the variety of shorter-timescale processes that could move a planetary system in AMD space. These typically occur on Myr timescales, during or immediately after formation. Processes linked to early formation, such as pebble accretion and migration through the gas disk, including migration trapping (e.g. \citealt{hansen_testing_2013, Moriarty16, dawson_correlations_2016, zawadzki_migration_2022}), cease far too early to explain a Gyr offset between ``multis" and "singles". Similarly, interactions with inclined and potentially eccentric distant giant companion that experienced an early epoch of scattering (e.g. \citealt{poon_origin_2020}, \citealt{pu_eccentricities_2018}, although other studies such as \cite{becker_effects_2017} found that distant perturbers typically were not sufficient to excite transiting planets out of transit), or excitation from a tilted, rapidly rotating host star, occur within the first 10-100 Myr \citep{spalding_spinorbit_2016, spalding_stellar_2020}. In the latter case, the quadrupole moment could exert a torque on the planetary orbits and dynamically heat the system. While such processes may be operative, they would change transit multiplicity over Myr rather than Gyr timescales \citep{chen_planets_2021}.  
 
 Several longer-timescale processes could hypothetically change transit multiplicity over Gyrs. \cite{pu_spacing_2015} first proposed a theoretical framework to explain the \textit{Kepler} transit multiplicity, wherein high-multiplicity systems (those with five or more transiting planets) are only metastable at birth. Subsequent dynamical instability events may, through collisions or ejections, lead to intrinsically fewer planets in systems (e.g. winnowing an original fully ``intact" population down to 20\% after a Gyr). These perturbations would also manifest in higher eccentricities and mutual inclinations on average, shifting planets out of transit geometry. Among other long-lived processes are the secular chaotic transfer of angular momentum deficit among the planets in a system into higher-eccentricity inner orbits, which would operate on a lower bound of 10-100 Myrs \citep{volk_dynamical_2020}, and secular chaotic diffusion due to overlapping three-body resonances \citep{petit_path_2020}. 

 We ultimately elect to adopt as agnostic a model as possible, requiring only of such a model that (1) some fraction of systems start ``intact" at birth (resembling the flat and densely populated systems that furnish the residual ``multis" population today), (2) dynamical temperature increases over time, (3) this probability of intactness decays as a power law, and (4) it hypothetically ceases at some point. Such a model can be described by a broken power law with four free parameters, that we collectively denote $\Theta$. These parameters govern the probability by which an originally ``intact" (STIP-type multi-transit) system becomes a ``disrupted" system. In this way, we assume very little about what physical mechanism drives dynamical excitation, in service of asking whether \textit{any} sculpting law capable of reproducing the \textit{Kepler} transit multiplicity could also produce a large age offset between singles and multis. 

These four free parameters, comprising $\Theta$, that describe the ``intact" probability as a function of stellar age are:
\begin{enumerate}
    \item The initial intact probability, assumed to be set at time $t=10^{8}$ yr. This is designated $p8$, ranging from 0 (meaning systems are all dynamically hot when 10$^{8}$ years have elapsed) to 1 (meaning all systems are STIPs at 10$^{8}$ years) in increments of 0.1. We sequester all dynamical sculpting information prior to $10^8$ years into this ``initial" intact probability, \init{}. 
    \item The decay rate from ``intact" to ``disrupted", $r$, necessarily $\le$0. The probability of intactness never increases as the star ages, but can only remain the same or decrease. We hereafter refer to \rate{} as the ``sculpting speed." We sample the sculpting speed between -1 to 0 dex$^{-1}$ in increments of 0.2.
    \item The stellar age $c$ at which dynamical evolution ceases and the intact probability stops decaying. We hereafter refer to this quantify as the ``cutoff time."  The cutoff time \cutoff{} varies from $10^8$ to $10^{10}$ years. By construction, a cutoff time \cutoff{} of $10^8$ years means no sculpting occurs in our model. A \cutoff{} of $10^{10}$ years, in contast, means that dynamical sculpting proceeds over cosmological timescales. We sample \cutoff{} at 11 points between 10$^{8}$ and 10$^{10}$ years in evenly log-spaced increments.
    \item The fraction of stars that host a planetary system, $f$. We sample $f$ from 0.1 to 1, in increments of 0.1.
\end{enumerate}

These sets of $\{p8, b, c, f\}$ are collectively our model parameters $\Theta$. We describe the form for our generalized sculpting equation here, as a function of $\Theta$. $P_{1}$ sets the probability of intactness at $10^{8}$ years, $P_{2}$ describes the era, between $10^{8}$ years and cutoff time $c$, during which ``intact" systems are actively decaying to ``disrupted", and $P_{3}$ describes the era when any sculpting has ceased. The probability function at all times is bounded between 0 and 1.

\begin{equation}
P_\textrm{intact}(\Theta) = \left\{
        \begin{array}{ll}
            P_{1,\textrm{intact}}(\Theta) & \quad \textrm{log}_{10}(\textrm{age}) \leq 8\\
            P_{2,\textrm{intact}}(\Theta) & \quad 8 < \textrm{log}_{10}(\textrm{age}) \leq \textrm{log}_{10}(\cutoff{}) \\
            P_{3,\textrm{intact}}(\Theta) & \quad \textrm{log}_{10}(\textrm{age}) > \textrm{log}_{10}(\cutoff{}),
        \end{array}
    \right.
\label{eq:p_intact}
\end{equation}

where 
\begin{equation}
    P_{1,\textrm{intact}}(\Theta) = p8
    \label{eq:p_intact1}
\end{equation}
\begin{equation}
    P_{2,\textrm{intact}}(\Theta) = p8 + r*(\textrm{log}_{10}(\textrm{age})-8)
    \label{eq:p_intact2}
\end{equation}
\begin{equation}
    P_{3,\textrm{intact}}(\Theta) = p8 + r*(\textrm{log}_{10}(c)-8).
    \label{eq:p_intact3}
\end{equation}

We employ the broken power law, generated as described above for each set of 4 free parameters $\Theta$, to generate the probability of ``intactness" independently to each star in our sample. Thus, based upon its age, each star receives a probability of hosting (1) an ``intact" system, (2) a ``disrupted" system, and (3) no planets. By randomly generating a uniform distribution from 0 to 1, we compare the random draw to the intact probability and assign its status accordingly. This ``intact" versus ``disrupted" status, in turn, determines how we assign and sample the multiplicity, orbital inclinations, and eccentricities of its planets.  We describe this procedure in Section \ref{sec:generating_planets}.

\subsection{Generating Planetary Systems}
\label{sec:generating_planets}
We aim ultimately to create a model transit multiplicity $\lambda$ that can be directly compared to the transit multiplicity from \textit{Kepler}. To achieve this end, we must first apply the sculpting law described by $\Theta$ to a set of host stars, populating those stars appropriately with planetary systems. We must then ``observe" the resulting sample to obtain a transit multiplicity yield. In this section, we describe the generation of planetary systems from $\Theta$. Given the dimensionality of our four-parameter grid: 11 values for initial intact fraction $p8$, 6 values for sculpting speed $r$, 11 values for age at which sculpting ceases $c$, and 10 values for the fraction of stars hosting planets $f$, this corresponds to 7260 unique $\Theta$. For each $\Theta$, we synthesize 30 unique stellar and planetary populations as follows.

Our stellar sample is drawn directly from the filtered cross-match described in Section \ref{sec:sample_selection}: stellar radii, masses, and 6-hour Combined Differential Photometric Precision (CDPP) are determined at a per-star level. Based on the stellar age and Equation \ref{eq:p_intact}, each system is assigned a probability $P_{\textrm{intact}}$. We draw a random number between 0 and 1, and by comparing each random draw to $P_{\textrm{intact}}$, the star is definitively assigned an ``intact" or ``disrupted" system of planets. 

The multiplicity, eccentricity, and orbital inclination distribution for each system depends on this assigned intact status. ``Intact" systems are assigned equal probability of having 5 or 6 planets, while ``disrupted" systems are assigned equal probability of having 1 or 2 planets \citep{ballard_thekeplerdichotomy_2016, zawadzki_migration_2022}. Midplanes for all systems and the longitudes of periastron for all planets are drawn uniformly from U(-$\pi$/2, $\pi$/2). We designate the randomly drawn midplanes as the zeros of mutual inclination; the zeros of longitude, meanwhile, are $\pi$/2. For ``intact" systems, mutual orbital inclinations are drawn from a Rayleigh distribution with width $\sigma$=2\degree, while mutual inclinations for planets in ``disrupted" systems are drawn from a Rayleigh distribution with width $\sigma$=8\degree. This is consistent with the intrinsic spread in $i$ required to reproduce the \textit{Kepler} transit multiplicity, whether modeled as one smooth population or two populations \citep{ballard_thekeplerdichotomy_2016, dawson_correlations_2016, zhu_about_2018, he_architectures_2020, zawadzki_migration_2022}. 

We next assign orbital eccentricity. Many studies have investigated the orbital eccentricities of exoplanet population, some specific to the \textit{Kepler} sample (e.g. \citealt{eylen_orbital_2019, Sagear23}). \citet{eylen_orbital_2019} proposed four distributions for sampling eccentricities: a) Rayleigh, b) half-Gaussian, c) Beta, and d) a mix of Rayleigh distributions for planets from disrupted systems and half-Gaussian distributions for planets from intact systems. We try these, as well as the eccentricity distribution from e) \citet{limbach_exoplanet_2015}, who fit eccentricities to the multiplicities of radial velocity exoplanets. All five heuristics yield higher eccentricities among dynamically hot systems than dynamically colder systems. To probe a more contrived scenario, we also try sampling eccentricities from a mixed model of the Rayleigh distribution for singles from \citet{eylen_orbital_2019} and the eccentricity distributions from \citet{limbach_exoplanet_2015} for multis -- this heuristic yields higher eccentricities among dynamically hot systems compared to the other eccentricity sampling heuristics. We ultimately elect to use a), a Rayleigh distribution with a peak at 0.24 for singles and a peak at 0.06 for multis \citep{eylen_orbital_2019}, finding that it offers the simplest heuristic while yielding similar results to even the contrived eccentricity sampling scheme. Finally, all planets are assigned the fiducial value of $R_{p}$= 2 $R_{\oplus}$.


\begin{figure}
\includegraphics[width=.42\textwidth]{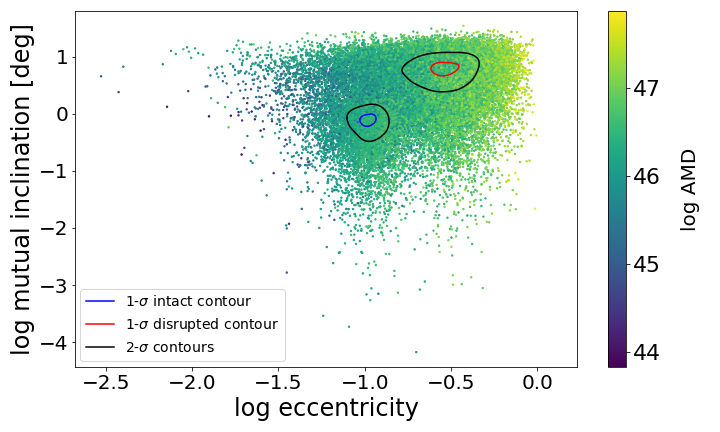}
\caption{Log angular momentum deficit (AMD) colored over the log eccentricity and mutual inclination distributions of simulated planetary systems based on one of our highest likelihood models. We note the trend of higher AMD with higher dynamical heat (that is, higher eccentricity and mutual inclination). Each dot represents the AMD and mean eccentricities and mutual inclinations of a system. The contours represent the 1- and 2-$\sigma$ populations for our simulated intact and disrupted systems (see Section \ref{sec:generating_planets}). }
\label{fig:amd_plot}
\end{figure}

\subsection{Calculating Transits and Completeness}
\label{sec:transits_and_completeness}

With the planet populations generated according to $\Theta$, we next determine which simulated planets transit their host star, based on the stellar radius, mass (to calculate the planet semi-major axis from Newton's version of Kepler's Third Law), and 6-hour CDPP. A planet geometrically transits if the absolute value of its impact parameter, calculated from Equation 7 in \citet{2010arXiv1001.2010W} as:
\begin{equation}
    b = \frac{a*\cos{i}}{R_*} \frac{1-e^2}{1+e*\sin{\omega}},
\label{eq:impact_parameter}
\end{equation}
is less than or equal to 1. In Equation \ref{eq:impact_parameter}, $a$ is the planet semi-major axis, $i$ is the planet's orbital inclination, $R_*$ is the stellar radius, $e$ is the planet's orbital eccentricity, and $\omega$ is the planet's longitude of periastron. Whether this geometric transit is necessarily observable also depends on the signal to noise ratio (SNR) of the transit. This is calculated using Equation 4 from \citet{2012PASP..124.1279C} as:
\begin{equation}
    \textrm{SNR} = \sqrt{\frac{t_{\textrm{obs}}\cdot f_0}{P}} \frac{(R_p/R_*)^2}{\textrm{CDPP}_{\textrm{eff}}}, 
\label{eq:snr}
\end{equation}

where
\begin{equation}
    \textrm{CDPP}_{\textrm{eff}} = \sqrt{\frac{t_{\textrm{CDPP}}}{t_{\textrm{dur}}}}{\textrm{CDPP}_N}.
\label{eq:cdpp_eff}
\end{equation}

Here, $P$ is the planet's orbital period, $R_p/R_*$ is the ratio of planet to stellar radius, $t\textsubscript{obs}$ is the time spanned observing the target (3.5 years), $f\textsubscript{0}$ is the fraction of that time spent actually observing, or duty cycle (0.92), t\textsubscript{CDPP} is 6 hours, and CDPP\textsubscript{N} is the rms CDPP of that duration. The transit duration is calculated based on Equations 14 and 16 from \citet{2010arXiv1001.2010W}:
\begin{equation}
    t_{\textrm{dur}} =  \frac{P}{\pi} \sin^{-1} {\frac{R_*}{a} \frac{\sqrt{(1+R_p/R_*)^2}-b^2}{\sin{i}}}.
\label{eq:t_dur}
\end{equation}

We then use the detection recovery rate ramp from Figure 2 of \citet{2013ApJ...766...81F} to calculate the probability of detection for a transiting exoplanet given its SNR:
\begin{equation}
P(\textrm{detection}) = \\
\left\{
        \begin{array}{ll}
            0 & \quad \textrm{SNR} \leq 6 \\
            0.1\times(\textrm{SNR}-6) & \quad 6 < \textrm{SNR} < 16 \\
            1 & \quad \textrm{SNR} \geq 16
        \end{array}
    \right.
\label{eq:fressin_detection}
\end{equation}

For each simulated, geometrically transiting planet in a system, we apply $P(\textrm{detection})$ to determine whether it is actually ``detected". For example, we should expect a system of two planets with P(detection) = [1.0, 0.7] to have one detected planet 30\% of the time and two detected planets 70\% of the time. Each system is therefore assigned a number of intrinsic planets and a number of ``detected" planets. With an ``observed" transit multiplicity per system, we tabulate the transit multiplicity distribution for that synthetic sample of stars. We now have in hand a set of transit multiplicity model yields (that is, a set of vectors $\lambda$ containing the number of systems with $n$ transiting planets, with $n$ ranging from 0 to 6) corresponding to each sculpting law $\Theta$ in our coarsely-sampled grid of free parameters. The range of $\lambda$ for each $\Theta$ incorporates both the random nature of transit geometry, and the likelihood of detection from the \textit{Kepler} sensitivity function. 

\subsection{Model Evaluation}
\label{sec:model_evaluation}


To evaluate the model parameters $\Theta$, we employ Bayes' theorem with a Poisson log likelihood as follows. In this expression, we quantify the probability of observing data $k$ (the transit multiplicity from \textit{Kepler}) given the sculpting parameters $\Theta$:
\begin{equation}
\begin{split}
    \textrm{log}(\mathcal{L}(\Theta|k)) =\textrm{log}(\mathcal{L}(k|\Theta)) = \\
    -\sum_{n=1}^{N} \ln(k!) - \sum_{n=1}^{N} \lambda + \sum_{n=1}^{N} (k*\ln(\lambda)),
    \end{split}
\label{eq:poisson_logL}
\end{equation}
where $\lambda$ is the simulated transit multiplicity generated from $\Theta$ and $N$ is the highest multiplicity. For bins with zero counts in them -- such as higher transit multiplicity bins for highly disrupted models -- the last term in Equation \ref{eq:poisson_logL} is set to zero, as done in \citet{vanhollebeke_stellar_2009}. 

\section{Results}
\label{sec:results}

\subsection{Injection and recovery}
\label{sec:injection-recovery}
\begin{figure*}[p]
  \sbox0{\begin{tabular}{@{}cccc@{}}
    \includegraphics[width=0.28\textwidth]{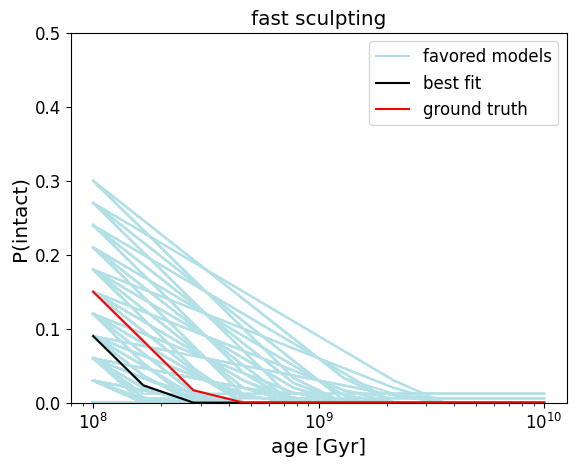}&
    \includegraphics[width=0.28\textwidth]{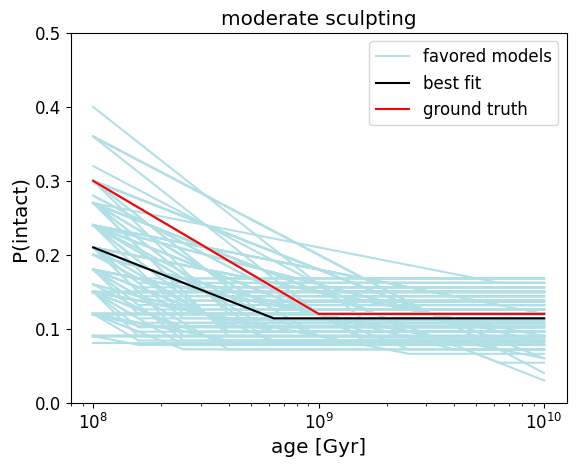}&
    \includegraphics[width=0.28\textwidth]{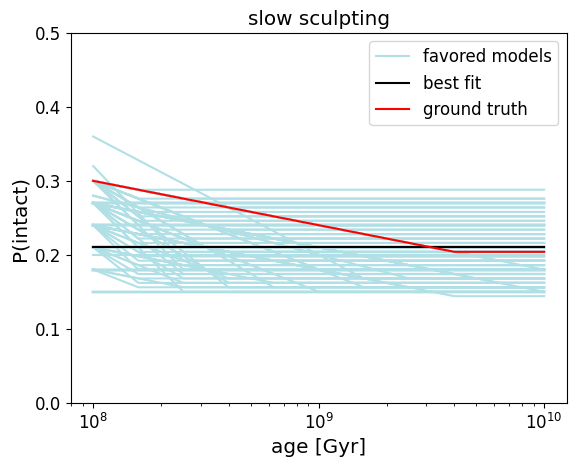}&
    \includegraphics[width=0.28\textwidth]{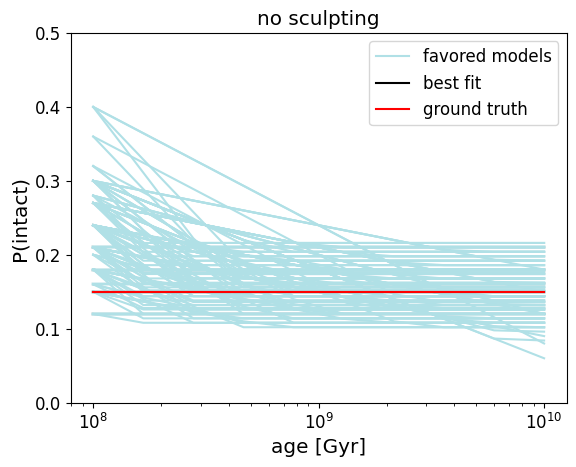}\\

\includegraphics[width=0.28\textwidth]{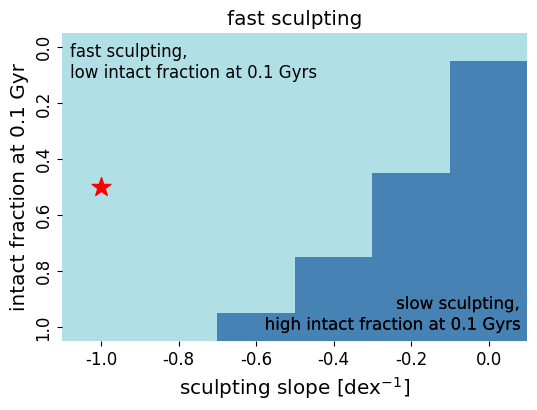} &
    \includegraphics[width=0.28\textwidth]{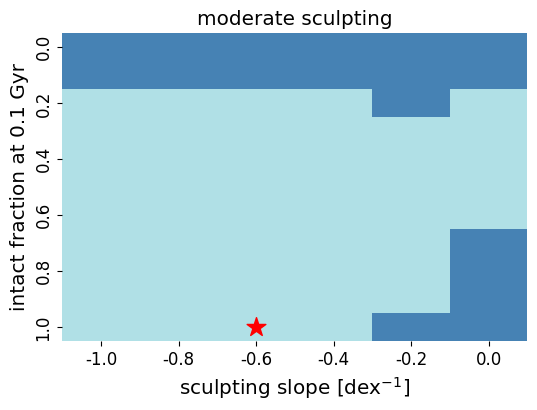}&
    \includegraphics[width=0.28\textwidth]{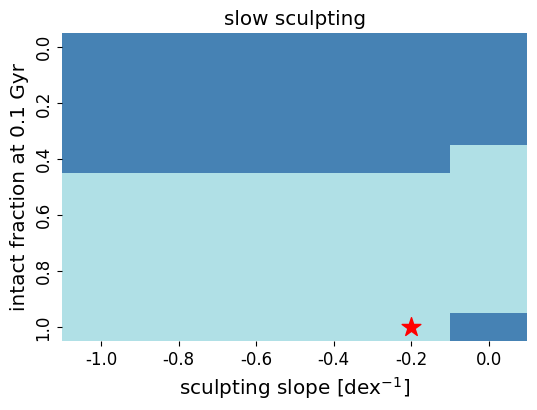} &
    \includegraphics[width=0.28\textwidth]{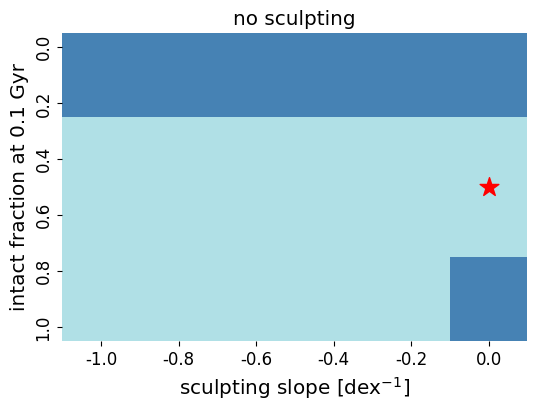} \\

  \includegraphics[width=0.28\textwidth]{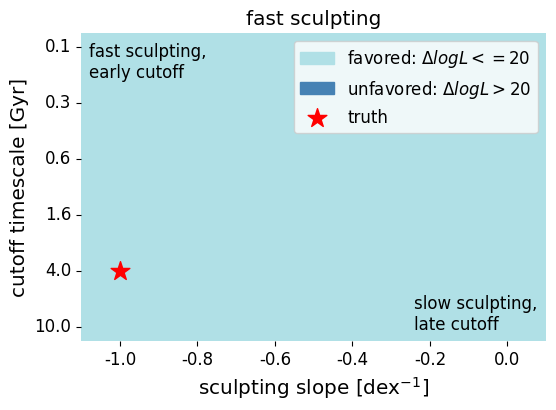}&
    \includegraphics[width=0.28\textwidth]{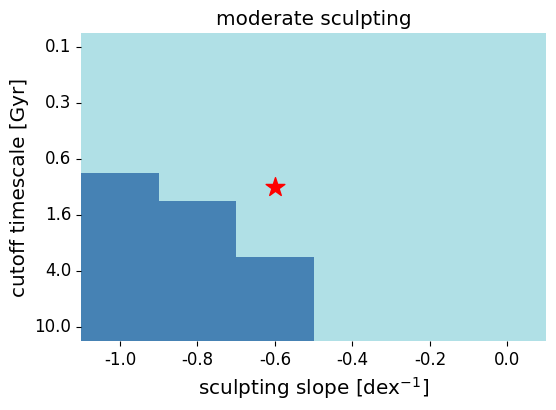}&
    \includegraphics[width=0.28\textwidth]{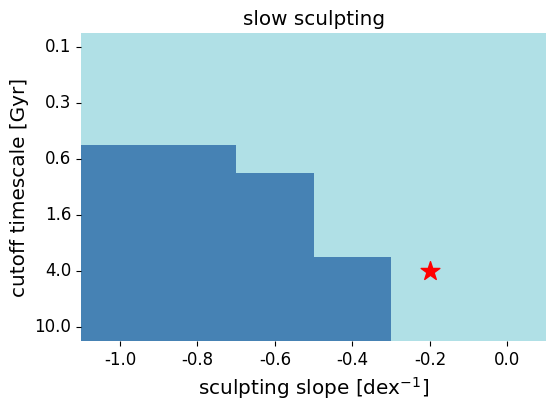}&
    \includegraphics[width=0.28\textwidth]{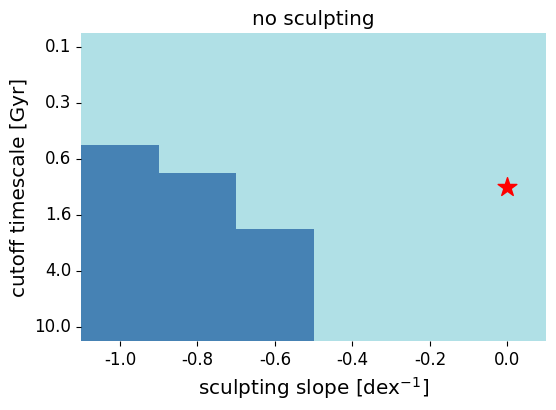}
  \end{tabular}}
  \rotatebox{270}{\begin{minipage}[c][\textwidth][c]{\wd0}
    \usebox0
    \caption{Constraints on sculpting parameters $\Theta$ resulting from four injection-recovery tests: fast (steep sculpting slope $r$), moderate, slow (shallow sculpting slope $r$), and no sculpting. \textbf{The top row} depicts the injected ground truth sculpting law in red, the recovered best-fit model in black, and all models with $\Delta\textrm{log}(\mathcal{L})<20$ of the best-fit model in blue. We note that in all injection-recovery tests, we recover a relatively tight range of present-day intact fractions, within 6\% of the ground truth. \textbf{The bottom two rows} depict two 2D slices of $\Theta$ parameter space from which the sculpting laws are constructed: the sculpting slope $r$ versus the intact fraction $p8$ at 10$^{8}$ yr, and the sculpting slope $r$ versus the sculpting cutoff timescale $c$. Pairs contributing to a ``favored" $\Theta$ are shaded in light blue, while darker blue regions correspond to pairs that do not appear in any ``favored" $\Theta$. We mark the ground truth $\Theta_{\textrm{true}}$'s location in these 2D slices with a red star, corresponding to the red sculpting law in the top row.}
  \label{fig:injection-recovery-fig}
  \end{minipage}}
\end{figure*}

We now perform a suite of injection-recovery tests, in order to evaluate our pipeline's sensitivity to the four free parameters that characterize our sculpting model. We simulate four ground-truth stellar and planet populations representing four different dynamical sculpting regimes with independent sets of $\Theta$. In each case, our planet host star fraction is set at 0.3. In Figure \ref{fig:injection-recovery-fig}, each sculpting law is depicted in red in the top row of figures. These injected $\Theta$ are:

\begin{enumerate}
    \item Scenario 1: ``fast sculpting" (\rate{} = -1, \init{} = 0.5, \cutoff{} = 4 Gyrs, \fraction{} = 0.3), resulting in a present-day intact fraction of 0\%. 
    \item Scenario 2: ``moderate sculpting" (\rate{} = -0.6, \init{} = 1.0, \cutoff{} = 1 Gyr, \fraction{} = 0.3), resulting in a present-day intact fraction of 12\%. 
    \item Scenario 3: ``slow sculpting"  (\rate{} = -0.2, \init{} = 1.0, \cutoff{} = 4 Gyrs, \fraction{} = 0.3), resulting in a present-day intact fraction of 21\%. 
    \item Scenario 4: no sculpting (\rate{} = 0., \init{} = 0.5, \fraction{} = 0.3; \cutoff{} is not applicable when there is no sculpting); this intact fraction is unchanging at 15\%. 
\end{enumerate} 

For each ground truth, we select one transit multiplicity yield $\lambda$ to represent the ``data". In generating these ground truth data sets, each star's age is known exactly to us: we draw these ages from the mean age values in the  \citet{berger_gaia-kepler_2020} sample. We use that age to assign per-system intact probability. Upon realistically ``observing" this sample, though, we must mimic having no exact knowledge of each stellar age. In this sense, our forward-modeling ``observation" pipeline operates in the same way, incorporating the stellar age uncertainty in \citet{berger_gaia-kepler_2020}. 

Then turning to our suite of models $\lambda$ generated from sculpting parameters $\Theta$, described in Section \ref{sec:generating_models}, we calculate the Poisson log likelihood for each $\Theta$, employing the injected ground truth as $k$ in Equation \ref{eq:poisson_logL}. Given the built-in variance associated with transit geometry and detection probability, we know that no model will exactly match the ``observed" $k$, including even the models drawn from the sculpting law that we know \textit{a priori} describes the data. Instead, we assess: (1) whether the ground truth sculpting model is in fact preferred by the data, and (2) to what extent other models perform equally well or better. 

We first report, that for all four ground truth cases, the range in $\textrm{log}\mathcal{L}$ between the ``observed" $k$, and the set of models $\lambda$ corresponding to the ground truth sculpting law, is $\le2$. We find as well that, among models for other sculpting laws, there often exists at least one $\lambda$ that performs equally well or better than the ground truth model, similarly furnishing a $\Delta\textrm{log}\mathcal{L}\le2$ of the ``best"-performing model. In practice, the range of $\Delta\textrm{log}\mathcal{L}$ among the $\lambda$ for these favored models will vary depending on the ground truth. We decide to designate ``favored" models as those with log($\mathcal{L})$ within 20 of the best-fit model, which is the threshold at which the picture of ``favored" and ``unfavored" models begins to stabilize across all four injection-recovery tests. This criterion includes all manifestations of any sculpting law that performs as well as the ground truth.

In Figure \ref{fig:injection-recovery-fig}, we depict the four injection-recovery tests and their results. The top row shows the injected ground truth sculpting model in red, the best-fit model in black, and models within 20 log($\mathcal{L})$ of the best-fit model in blue. In the bottom two rows of Figure \ref{fig:injection-recovery-fig}, we depict two slices of the parameter space from which we build our sculpting laws: \rate{} versus \init{}, and \rate{} vs \cutoff{}. The red star indicates the location of the injected ground truth sculpting law within parameter space. Light blue parameter space corresponds to pairs of free parameters that produce ``favored" models, while dark blue corresponds to pairs of parameters for which no $\lambda$ furnishes a log($\mathcal{L})$ within 20 of the best-fit model. We annotate the corners of these plots to indicate the general behavior of sculpting laws in that region (e.g. ``fast", ``late cutoff").

We identify several trends in the degeneracies between parameters, and in our ability to rule out $\Theta$ that are inconsistent with the data. 

\begin{enumerate}
    \item In each case, the $\Theta_{\textrm{true}}$ governing that sample always lies within the set of $\{\Theta\}$ corresponding to ``favored" sculpting laws. In this sense, we successfully recover the injected dynamical sculpting parameters. However, there is no instance in which we can uniquely recover $\Theta_{\textrm{true}}$. Rather, there are other regions of parameter space that are also ``favored" per our $\Delta\textrm{log}\mathcal{L}\le20$ criterion above. By design (given that we evaluate likelihood based upon transit multiplicity alone), we cannot distinguish between $\Theta_{\textrm{true}}$ and a set of parameters $\Theta$ that similarly reproduces the present-day ``intact fraction" of planetary systems. 

    \item Our ability to reverse-engineer $\Theta_{\textrm{true}}$, then, is set by our precision in recovering the present-day intact fraction. In all cases, the sets of parameters $\{\Theta\}$ resulting in favored models generated intact fractions within $\sim6$\% of the truth. For Scenario 1, we constrain the intact fraction to be  0-2\% (ground truth of 0\%), for Scenario 2 we constrain it to be 7-18\% (ground truth of 12\%), for Scenario 3 we constrain it to be 15-27\% (ground truth of 21\%) and for Scenario 4 we constrain it to be 10-21\% (ground truth of 15\%). Any $\Theta$ that produces a present-day intact fraction within 6\% of the true intact fraction is favored, regardless of whether $\Theta_{\textrm{true}}$ involved ``fast", ``moderate" or ``slow" sculpting speeds. Taking Scenario 2, ``moderate sculpting", as a case study (21\% intact today), we cannot distinguish between limiting cases in which (1) 100\% of planet-hosting systems are intact at 10$^{8}$ years, decaying to only 50\% likelihood of intactness at 1 Gyr (corresponding to $\Theta_{\textrm{true}}$), and another case in which (2) 50\% of planet-hosting systems are intact at 10$^{8}$ years, and no sculpting occurs whatsoever after that point.

    \item Conversely, any set of $\Theta$ that \textrm{cannot} produce this resulting intact fraction are disfavored. In this sense, we are able to identify sculpting laws that are \textit{inconsistent} with the transit multiplicity today. These $\Theta$ fall into two categories: those that result in too many ``multis", and those that result in too few. For the same ``moderate" sculpting case (producing a present intact fraction of 12\%), we rule out the following illustrative $\Theta$:
    \begin{itemize}
    \item \textbf{Example sculpting laws $\Theta$ disfavored for producing too \textit{few} multis}: Any rapid decay of -1.0 dex $^{-1}$ from intact to disrupted proceeding for longer than 600 Myr, any moderate decay of -0.6 dex $^{-1}$ proceeding for $\ge$4 Gyr, or any $\Theta$ with initial intact fraction of systems at 10$^{8}$ $\le$0.1. 
    
    \item \textbf{Example sculpting laws $\Theta$ disfavored for producing too \textit{many} multis}: models with no decay, if $\ge$60\% of systems are intact at 10$^{8}$ years. In this case, too \textit{many} intact systems would remain, higher than the highest allowable intact fraction consistent with the transit multiplicity (18\%). 
    \end{itemize}

\end{enumerate}

For our ``fast" sculpting scenario, in which 0\% of systems remain intact today, we rule out only $\Theta$ that furnish too many multis. Initial intact fractions at 10$^{8}$ years $>0$ are consistent only if decay occurs rapidly in a commensurate way: e.g. if 100\% of systems start intact, then intact probability must decay as -1 or -0.8 dex$^{-1}$. Otherwise, too many intact systems would remain in the sample to be consistent with the observed transit multiplicity. Conversely, if the intact fraction at 10$^{8}$ years is sufficiently low (close to 0\% to begin with), then all subsequent sculpting laws are permissible (depicted by the fact, in the bottom row of Figure \ref{fig:injection-recovery-fig}, that all sculpting speeds and cutoff timescales are permissible for at least one $\Theta$).

\subsection{Application to \textit{Kepler} sample}
\label{sec:application_to_kepler}

After assessing our sensitivity to sculpting laws via injection-and-recovery, we proceed to characterize $P(k|\Theta)$, where $k$ is the real \textit{Kepler} transit multiplicity (Section \ref{sec:model_evaluation}). As described above in Section \ref{sec:generating_models}, we have in hand a suite of model transit multiplicities $\lambda$ generated according to a grid of possible sculpting laws $\Theta$: 30 for each $\Theta$ (the range in these 30 $\lambda$ encompasses variance from intact probability, transit geometry, and detection likelihood). We define our range of acceptable $\Theta$ as those that furnish a $\Delta\textrm{log}(\mathcal{L})<20$ of the highest log likelihood. This criterion is motivated by our findings from the injection-and-recovery analysis described in \ref{sec:injection-recovery}.

\begin{figure}
\includegraphics[width=.45\textwidth]{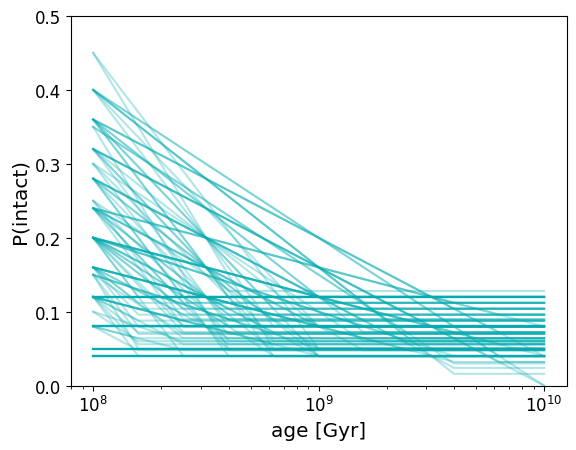}
\includegraphics[width=.45\textwidth]{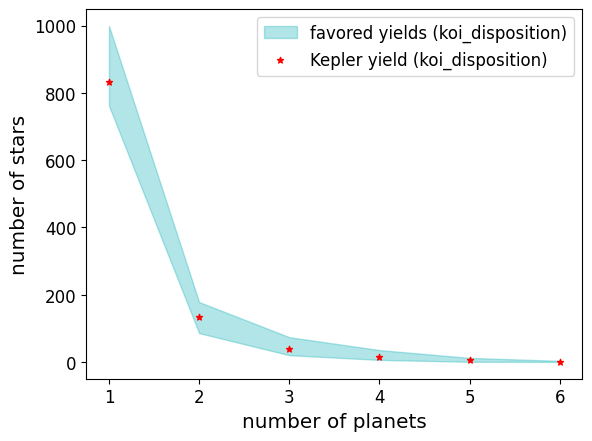}
\caption{\textbf{Top:} We show sculpting laws with log($\mathcal{L})$ $\geq$ -34.5, or the best log($\mathcal{L})$ minus 20. These models represent <3\% of the models we run through the forward modeling pipeline. 
\textbf{Bottom:} Here, we show the transit multiplicity yields for these favored models. The blue envelope illustrates the full range of these favored model yields for each transit multiplicity bin, and the red stars represent the {\it Kepler} transit multiplicity for FGK dwarfs, using the koi\_disposition planet vetting criterion.}
\label{fig:best_models}
\end{figure}

The top panel of Figure \ref{fig:best_models} shows all sculpting laws $\Theta$ for which at least one $\lambda$ furnishes a log likelihood within 20 of the best log($\mathcal{L})$ across all models (this highest log likelihood is -14.5). The bottom panel of Figure \ref{fig:best_models} shows the transit multiplicity yields of these favored dynamical sculpting models, together with the observed transit multiplicity based on the koi\_score planet vetting criterion, shown in red symbols. These favored models, whose transit multiplicities match well with the observed {\it Kepler} transit multiplicity, represent less than 3\% of the models we pass through our forward modeling pipeline. 

As in the injection-and-recovery tests described in Section \ref{sec:injection-recovery}, we depict 2D slices of $\Theta$, indicating in light blue pairs of parameters that correspond to at least one ``favorable" sculpting law (log($\mathcal{L})$ $\geq$ -34.5) and in darker blue pairs of parameters for which no $\Theta$ furnishes a likelihood above this criterion. The top row corresponds to the 2D slices we showed in Figure \ref{fig:injection-recovery-fig}: sculpting slope versus intact fraction at 0.1 Gyr, and sculpting slope versus the cutoff timescale. 

\begin{figure*}
  {\begin{tabular}{@{}cc@{}}
    \includegraphics[width=0.45\textwidth]{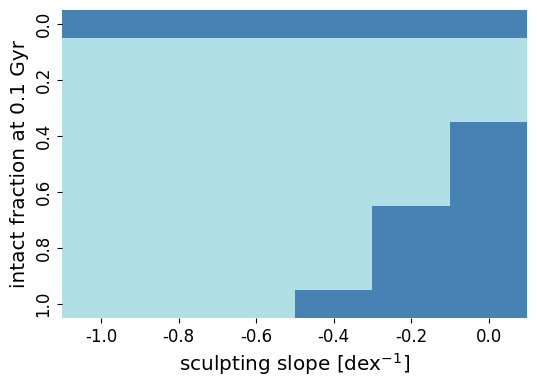}&
    \includegraphics[width=0.45\textwidth]{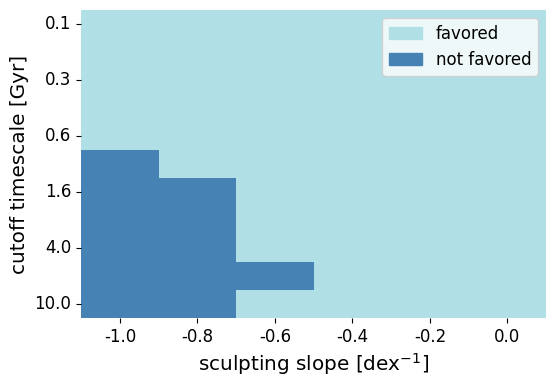}\\
    \includegraphics[width=0.45\textwidth]{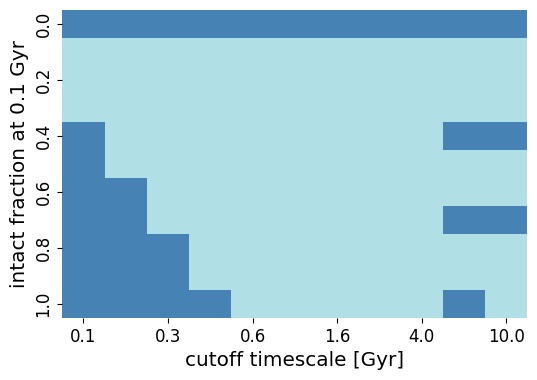}&
    \includegraphics[width=0.45\textwidth]{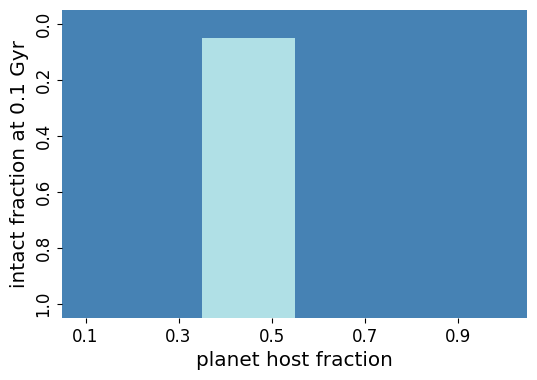}

  \end{tabular}}
    \caption{We show four different 2D slices of the free parameter space, where each cell is colored in light blue if the best sculpting law in that 2-tuple group of models has a log likelihood greater than or equal to -34.5 and dark blue otherwise. \textbf{Top left:} Our forward modeling pipeline disfavors sculpting laws that have shallow or zero slope, unless they are sufficiently compensated by a low intact fraction at 0.1 Gyr. \init{} of 0 is categorically disfavored. \textbf{Top right:} Sculpting laws that have steep slopes and cutoff timescales beyond 0.6 Gyr are disfavored. \textbf{Bottom left:} We disfavor sculpting laws that have cutoff timescales shorter than 0.6 Gyr and an intact fraction at 0.1 Gyr that is insufficiently low to offset the overproduction of multis attendant with short-lived sculpting. \textbf{Bottom right:} We appear to favor a planet host fraction of 0.4 and 0.5.}
  \label{fig:results}
\end{figure*}

As before, all $\Theta$ that reproduce the present-day intact fraction with a given precision are statistically indistinguishable. The family of $\Theta$ that are unlikely fall in two categories: those that produce too many ``multis", and those that produce too little. 
Looking at the 2D projection of sculpting slope versus intact fraction (top left panel of Figure \ref{fig:results}), we can rule out shallow sculpting laws that, combined with moderate to high intact fractions at 0.1 Gyr, result in an overabundance of multis. This is consistent with the scenario of a fast sculpting ground truth from the injection-recovery tests, although slower sculpting ground truths have this relation to a less pronounced extent. The 2D projection of sculpting slope versus cutoff timescale (top right panel of Figure \ref{fig:results}) presents the opposite case: sculpting laws that are both steep and level off too late result in too much dynamical sculpting, leading to fewer multis than observed by \textit{Kepler}. This latter case is apparent in the moderate, slow, and no-sculpting injection-recovery test cases. We also rule out \init{}s of zero, which is not a feature of the fast sculpting case but is apparent in the other cases. This tension may indicate that a universal sculpting law is in reality somewhere between the fast and moderate cases, or that there exist multiple possible sculpting laws among \textit{Kepler} systems, with some faster than others.

When we consider the 2D slice of sculpting cutoff timescale versus the intact fraction at 0.1 Gyr (bottom left panel of Figure \ref{fig:results}), we can rule out short-lived sculpting processes that end between 0.1 and 0.4 Gyr, unless they are compensated by sufficiently low \init{}. This feature is common among all four injection-recovery cases explored in Section \ref{sec:injection-recovery}, although the extent of this region of disfavored models varies from case to case. Finally, the bottom right panel of Figure \ref{fig:results} groups models by planet host fraction and the intact fraction at 0.1 Gyr. This panel suggests that the \textit{Kepler} transit multiplicity favors a planet host fraction among Sun-like \textit{Kepler} stars of 0.4 or 0.5. Our forward modeling pipeline's strong favoring of these two discrete planet host fractions is a reflection of the coarseness of our grid in $\Theta$.

Our range of favored planet host fractions lies within or close to the range of values uncovered by other authors. It is somewhat higher than the findings of \cite{zhu_about_2018}, who found that 30\% $\pm$ 3\% of Sun-like stars have {\it Kepler}-like planetary systems, with planetary radii greater than that of the Earth and orbital periods less than 400 days. It is spanned by the 0.56$^{+0.18}_{-0.15}$ {\it Kepler}-like planet host fraction derived by \citet{he_architectures_2019}, and it includes the 0.441$\pm$0.019 planet host fraction found by \cite{bashi_exoplanets_2022}. 

\subsection{Fraction of intact systems}
\label{sec:intact_frac}
\begin{figure}
\includegraphics[width=.45\textwidth]{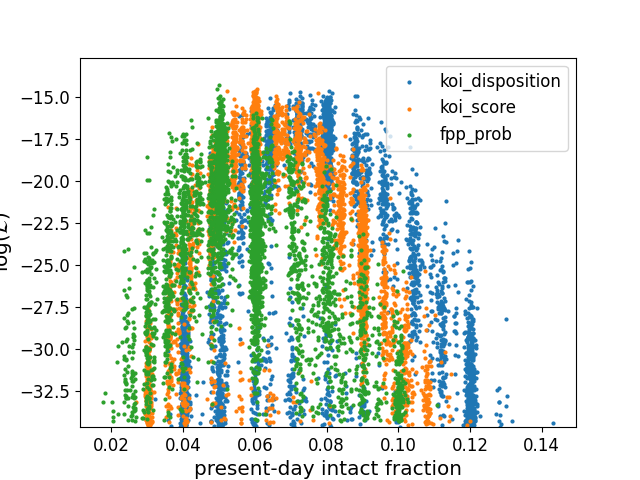}
\caption{We plot favored models (those models with log likelihood within 10 of the maximum likelihood model) as a function of the present-day intact fractions from which their planet populations were drawn. Each dot represents a realization of a sculpting model's fraction of \textit{Kepler} FGK dwarfs that host STIPs and the associated log likelihood. Using the koi\_disposition indicator, the intact fraction peaks in likelihood at 8\%, and the present-day intact fraction among favored models ranges from 4\% to 13\%. We show that our models are robust to the choice of KOI indicator, with the favored intact fractions overlapping between all three indicators used.}
\label{fig:intact_frac}
\end{figure}

As we uncovered from our suite of injection-and-recovery tests, the singular criterion that signifies that a sculpting law $\Theta$ is consistent with the data is its ability to correctly produce the fraction of ``intact" systems today. Figure \ref{fig:intact_frac} plots the favored models against the present-day intact fractions from which their planet populations are produced. Each dot represents a realization of a model (some favored models may not have all 30 of their realizations satisfy the criterion of $\textrm{log}(\mathcal{L})$ $\geq$ -34.5). For all three KOI disposition indicators, the favored present-day intact fraction is relatively consistent. Using the koi\_disposition indicator, we find that the peak fraction of \textit{Kepler} FGK stars that host dynamically cool STIPs is 8\%, and -- among the favored models -- this fraction ranges from 4\% to 13\%. Dividing this by the fraction of planet hosts yields the fraction of planet-hosting FGK dwarfs whose systems are dynamically cool. 

From our favored planet-host fractions of 0.4 and 0.5, we therefore infer that 8-33\% of planet-hosting \textit{Kepler} FGK stars host systems of tightly-packed inner planets (STIPs). This range is lower than the result from \citet{mulders_exoplanet_2018}, which found that at least 42\% of FGK dwarfs have nearly coplanar planetary systems with at least seven planets. Instead, our findings adhere much more closely to the result of \citet{lissauer_architecture_2011}, who found that approximately 3\%–5\% of Kepler target stars host a ``STIP" with multiple planets with radii between 1.5 $R_{\Earth}$ and 6 $R_{\Earth}$ and orbital periods between 3 and 125 days. We also find that our favored models correspond to a present-day fraction of dynamically hot planetary systems ranging from 0.26 to 0.46.

\subsection{Transit multiplicity versus stellar age}
\label{sec:alternatives}

Given the set of dynamical sculpting laws $\Theta$ that are consistent with the \textit{Kepler} transit yield, we can predict the corresponding behavior of other observables. Specifically, we consider here how the relationship between stellar age versus transit multiplicity $n$ ought to behave. We can then test that predicted behavior for its consistency with previously observed trends. 

\begin{figure}
\includegraphics[width=.45\textwidth]{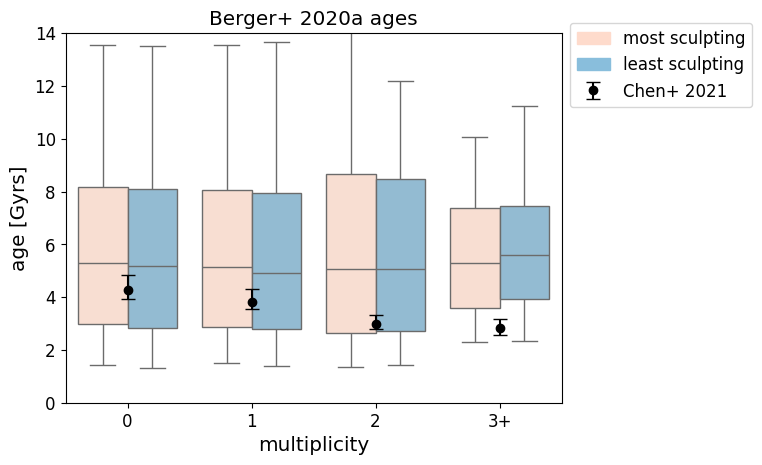}
\includegraphics[width=.45\textwidth]{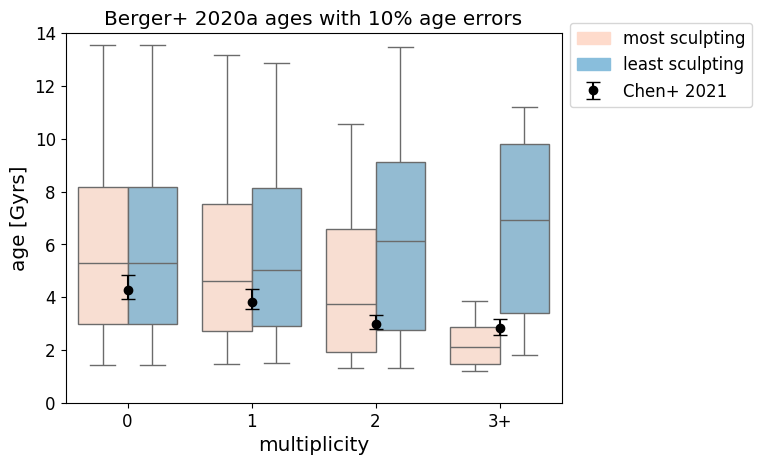}
\caption{\textbf{Top:} We compare the age-multiplicity relations of two favored models to that of \citet{chen_planets_2021}. The two models were chosen such that one has the greatest change in $P_{intact}$ over the relevant timescale (pink; ``most sculpting") and the other has the least change in $P_{intact}$ over the same timescale (blue; ``least sculpting"). The boxes contain the middle 50\% age uncertainties; the whiskers denote the 5th and 95th age percentiles. Using our sample, we find no significant change in multiplicity with age. \textbf{Bottom:} We conduct the same comparison running the same models through a stellar sample redrawn using 10\% age errors, showing that it requires an idealized scenario of exquisitely measured stellar ages to produce a better match to the \citet{chen_planets_2021} result.}
\label{fig:boxplot}
\end{figure}

We employ the same set of favored $\Theta$ shown in the top panel of Figure \ref{fig:best_models}. As described in Section \ref{sec:generating_planets}, each $\Theta$ corresponds to an independent draw of the planetary systems populating the \textit{Kepler} stars (for which ages are drawn from \citealt{berger_gaia-kepler_2020}). We repeat the sample generation 30 times for each dynamical sculpting law, before investigating the resulting detected transit multiplicity for the sample, $\lambda$. We consider, therefore, the $\sim$3\% of synthetic stellar samples that reproduce the \textit{Kepler} transit multiplicity, tallying now the stellar ages associated with hosts to \textbf{no transiting planet}, 1 transiting planet, 2 transiting planets, and so on. Our findings are shown in Figure \ref{fig:boxplot}, where the pink boxplot represents the age distribution for each transit multiplicity bin given the favored sculpting law that results in the greatest change in $P_{intact}$, and the blue boxplot represents the same for the favored sculpting law that results in the least change in $P_{intact}$. 

Within the set of all $\Theta$, there \textit{are} sculpting laws that persist over $Gyr$ timescales and are hypothetically able to reproduce the 1.5 Gyr offset observed by \cite{chen_planets_2021}. Yet, in practice we find that no \textit{favored} sculpting law produces this finding: in other words, no $\Theta$ that matches the \textit{Kepler} transit multiplicity also predicts an observably significant change in transit multiplicity with age in the \textit{Kepler} sample (Figure \ref{fig:boxplot}).

We understand this finding in the context of two features of our sample, one that is due to our stellar age uncertainties, and one that is intrinsic to the nature of the sculpting laws we consider. Firstly (1) our age uncertainties are much larger than those of \citet{chen_planets_2021}. While they report an age of field stars ${4.28}_{-0.35}^{+0.55}$ Gyr, we find that hosts to no transiting planets exhibit a much wider spread in ages at the 1$\sigma$ level. Using the isochrone ages from \cite{berger_gaia-kepler_2020}, 68\% of our stars have ages within $5.2_{-2.2}^{+2.8}$ Gyr. While this range is consistent with the \cite{chen_planets_2021} age range, our stellar age uncertainty is about a factor of 5 higher. Even though we do, in fact, see a mean offset of 1.5 Gyr between hosts to 0 transiting planets and hosts to $\ge$3 transiting planets, the spread in ages is so large as to render the offset statistically insignificant. 

However, another reason for the ``washing out" of the multiplicity versus stellar age relationship is attributable to the inevitable sample contamination between ``singles" and ``multis", and indeed between field stars and planet hosts. \textit{Bona fide} dynamically cold systems, even if they are offset significantly in age, will not only populate the 3+ transiting planet sample. Referring back to the high- and low-sculpting favored models represented in Figure \ref{fig:boxplot}, when we calculate the fraction of single-transiting systems, double-transiting systems, and so on that come from intact versus disrupted systems, we find a relatively consistent 16-17\% of the sample of singles is attributable to dynamically cold ``intact" systems for which only one planet transits. Similarly, dynamically hot systems will occasionally furnish more than one transiting planet, contaminating the sample of ``multis." 

We summarize these statistics in Table \ref{table1}, which shows intact and disrupted system membership for different transit multiplicity bins for the two favored models (where log($\mathcal{L})$ $\geq$ -34.5) from Figure \ref{fig:boxplot}: one exhibiting the greatest change in $P_{intact}$ over the relevant timescale (``most sculpting") and the other exhibiting the least change in $P_{intact}$ (``least sculpting"). Multis here refer to systems with 3 or more transiting planets, following the definition used by \citet{brewer_compact_2018}.

\begin{table}
  \centering
    \begin{tabular}{ c|c|c }
    Model \& & Intact & Disrupted \\
    multiplicity & membership & membership \\
     \hline
     Most sculpting, singles & 0.17 & 0.83 \\  
     Most sculpting, doubles & 0.64 & 0.36 \\  
     Most sculpting, multis & 1.00 & 0.00 \\  
     \hline
     Least sculpting, singles & 0.16 & 0.84 \\
     Least sculpting, doubles & 0.70 & 0.30 \\
     Least sculpting, multis & 1.00 & 0.00 \\
    \end{tabular}  \caption{Intact and disrupted system membership, among favored models}
  \label{table1}
\end{table}

\subsection{Prospects for constraints with more precise age measurements}
\label{sec:strong_age_constraint}

Our ability to infer dynamical sculpting timescales hinges critically on our ability to estimate stellar ages: indeed, our larger uncertainties in stellar age account for one of the reasons why we cannot reproduce a detectable 1.5 Gyr offset between systems with 0 transiting planets and those with several. We aim in this Section to quantify our ability to constrain the dynamical sculpting law $\Theta$ in ideal conditions: with very strong age constraints in hand. Hypothetically, stellar age measurements at the fidelity of asteroseismic ages -- but for all stars -- would improve our ability to recover dynamical sculpting laws at the timescales we specifically investigate in this work. 

\begin{figure*}[p]
  \sbox0{\begin{tabular}{@{}cccc@{}}
    \includegraphics[width=0.28\textwidth]{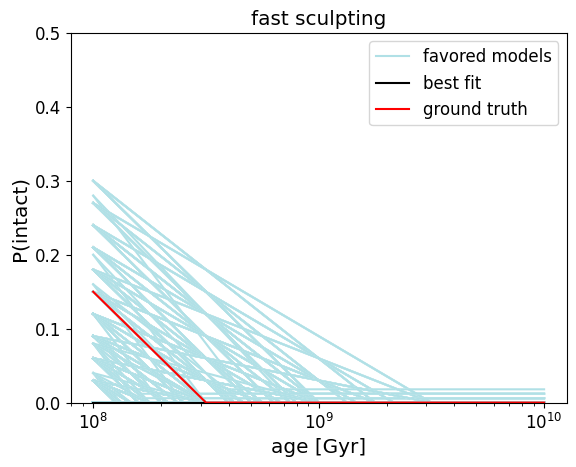}&
    \includegraphics[width=0.28\textwidth]{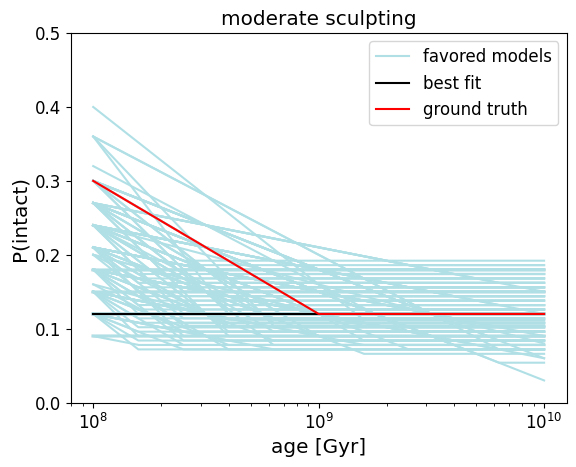}&
    \includegraphics[width=0.28\textwidth]{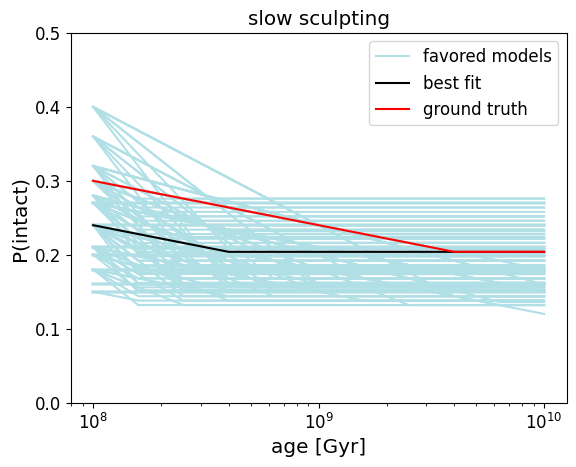}&
    \includegraphics[width=0.28\textwidth]{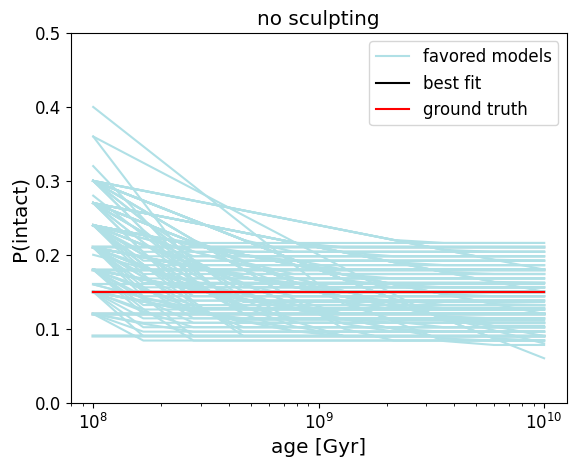}\\

\includegraphics[width=0.28\textwidth]{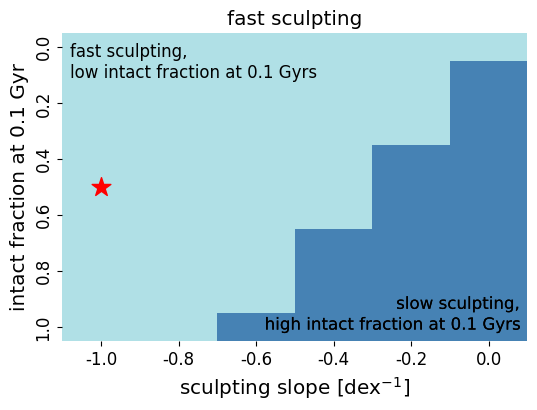} &
    \includegraphics[width=0.28\textwidth]{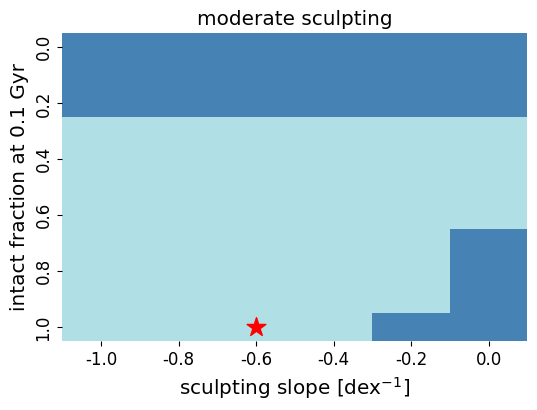}&
    \includegraphics[width=0.28\textwidth]{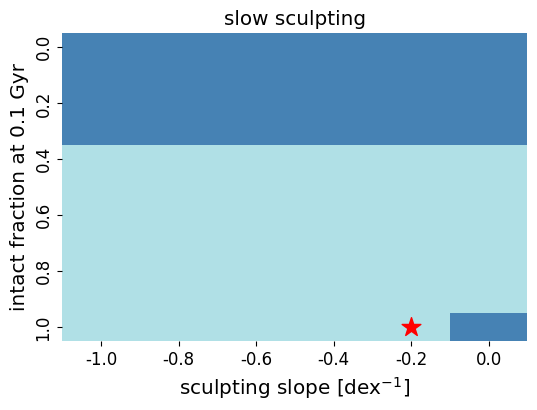} &
    \includegraphics[width=0.28\textwidth]{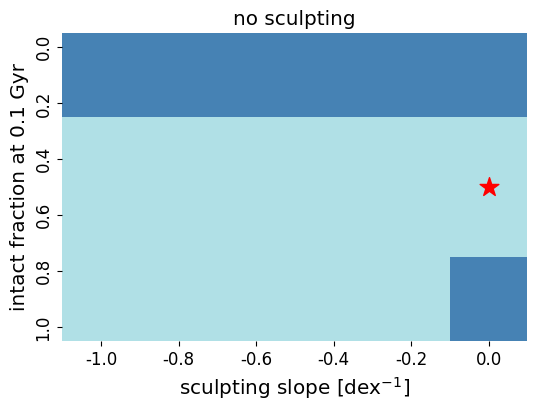} \\

  \includegraphics[width=0.28\textwidth]{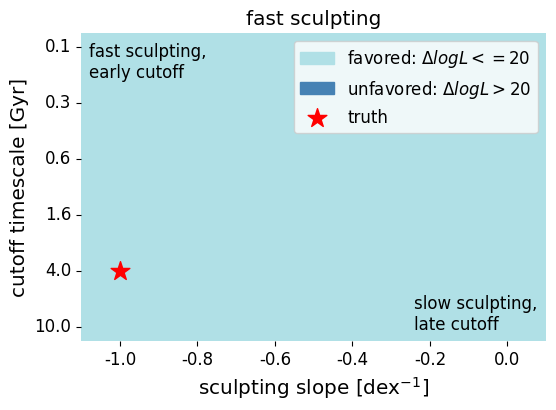}&
    \includegraphics[width=0.28\textwidth]{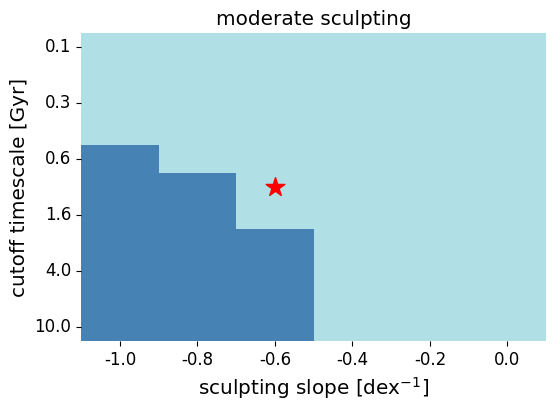}&
    \includegraphics[width=0.28\textwidth]{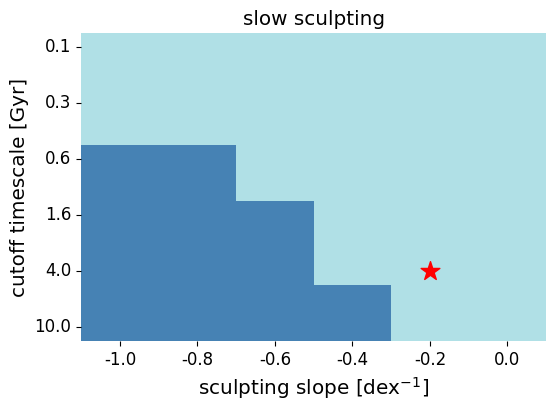}&
    \includegraphics[width=0.28\textwidth]{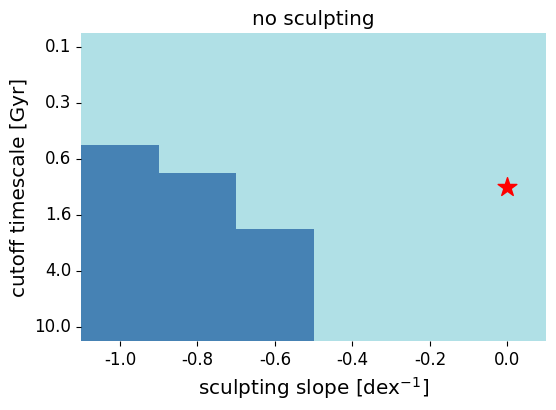}
  \end{tabular}}
  \rotatebox{270}{\begin{minipage}[c][\textwidth][c]{\wd0}
    \usebox0
    \caption{We show the same series of four injection-recovery tests depicted in Figure \ref{fig:injection-recovery-fig}, but with uniformly distributed 10\% age errors, instead of those derived from isochrones and provided in the \citet{berger_gaia-kepler_2020} \textit{Gaia}-\textit{Kepler} cross-match. We find that, besides a slight improvement in the ability to rule out marginally higher \init{}, there is no significant difference between what we can recover with our current age estimates and what we can recover in the idealized case of 10\% age errors.}
    \label{fig:injection-recovery-ten}
  \end{minipage}}
\end{figure*}

We run the same injection-recovery tests as in Section \ref{sec:injection-recovery}, with an important change: the stellar sample we populate with planetary systems now has a constant 10\% age uncertainty for all stars. We show the results of these tests in Figure \ref{fig:injection-recovery-ten}. Besides some increased ability to rule out slightly higher \init{}, there exists no significant difference in our ability to recover the injected $\Theta$. The lack of difference between these two batteries of injection-recovery tests suggests an upper limit of what we can infer about dynamical sculpting timescales from the \textit{Kepler} transit multiplicity: there is so much intra-sample contamination between ``singles" and ``multis" that dynamical sculpting alone cannot explain an offset between them. We can rule out large swaths of parameter space for being inconsistent with the transit yield, but there exist a wide range of $\Theta$ that furnish equally likely models. 

While painting 10\% age errors onto our sample does little to improve our ability to forward model the possible regimes of dynamical sculpting among many $\Theta$, the bottom panel of Figure \ref{fig:boxplot} suggests that it is possible in this idealized case for at least one model to produce a noticeable inverse relation between age and multiplicity. We conclude that \textit{current stellar age measurements are insufficient to support any dynamical sculpting law $\Theta$ that can produce a significant and detectable offset in ages of multis versus singles; however, better stellar age constraints may enable us to uncover such an offset.}

\subsection{Prospects for constraints with changes to stellar age sample}
\label{sec:ages}

\begin{figure*}[p]
  \sbox0{\begin{tabular}{@{}cccc@{}}
    \includegraphics[width=0.28\textwidth]{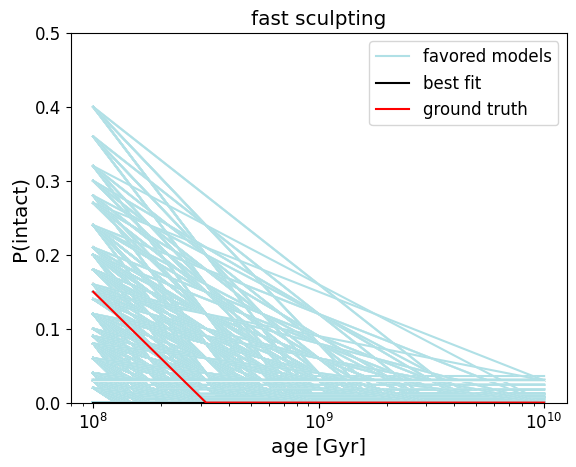}&
    \includegraphics[width=0.28\textwidth]{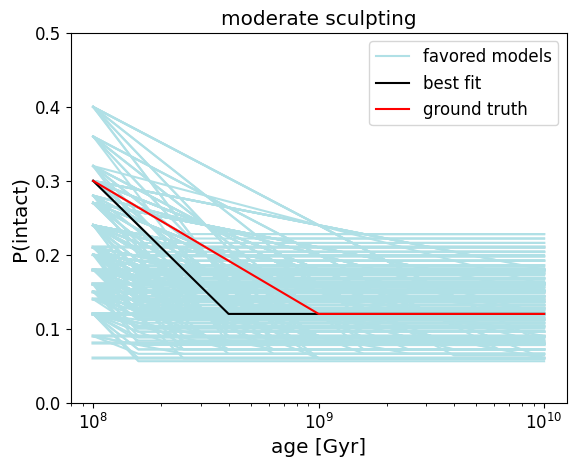}&
    \includegraphics[width=0.28\textwidth]{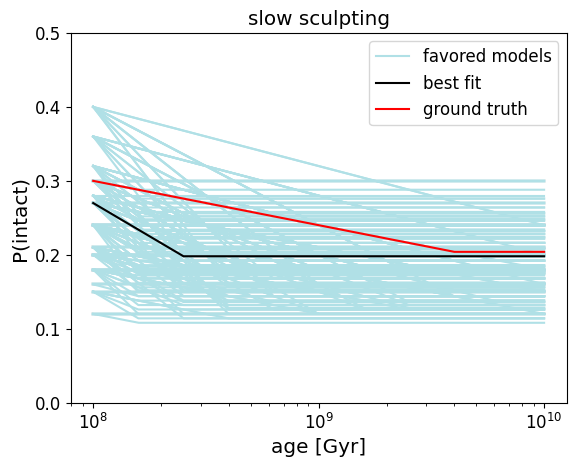}&
    \includegraphics[width=0.28\textwidth]{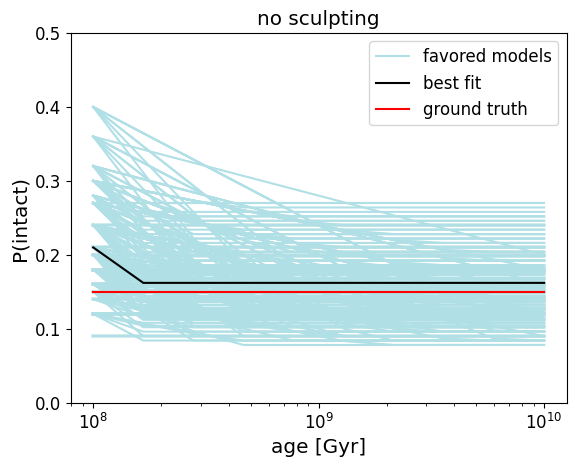}\\

\includegraphics[width=0.28\textwidth]{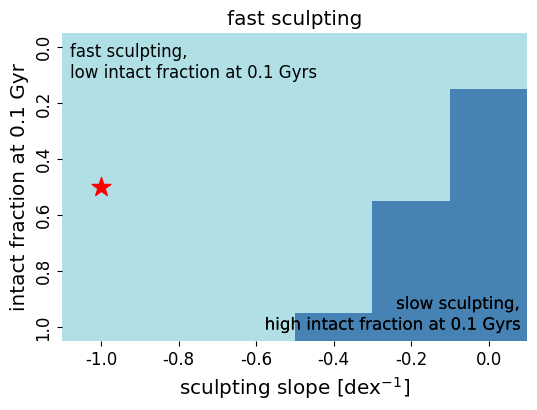} &
    \includegraphics[width=0.28\textwidth]{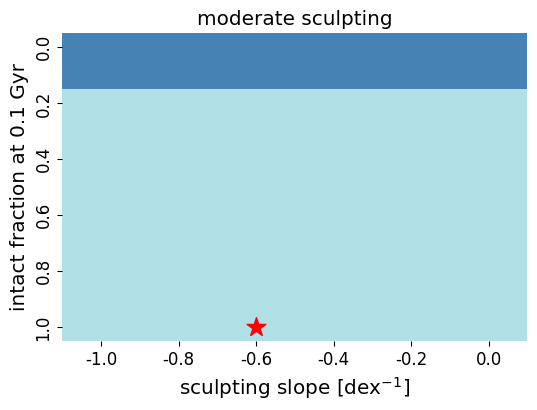}&
    \includegraphics[width=0.28\textwidth]{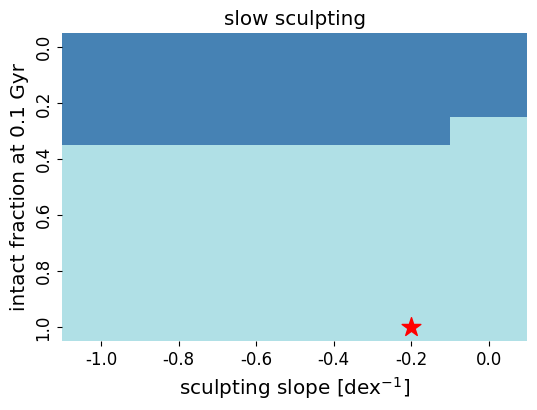} &
    \includegraphics[width=0.28\textwidth]{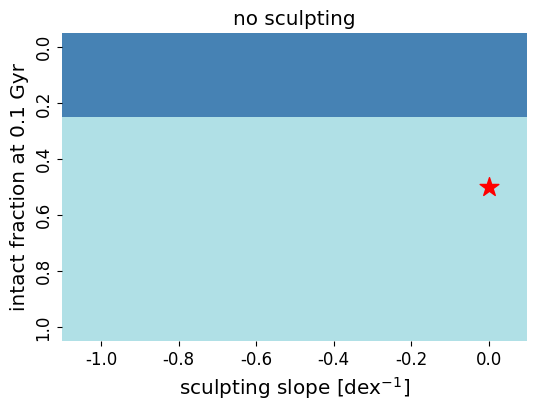} \\

  \includegraphics[width=0.28\textwidth]{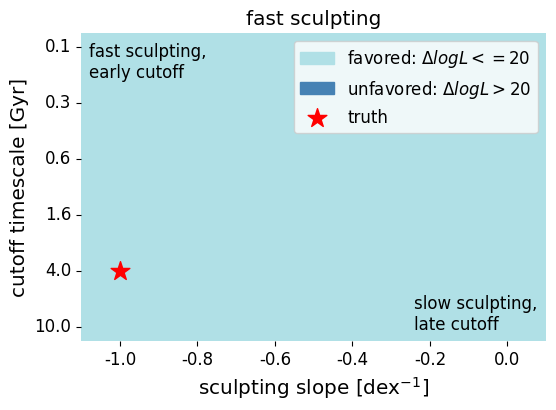}&
    \includegraphics[width=0.28\textwidth]{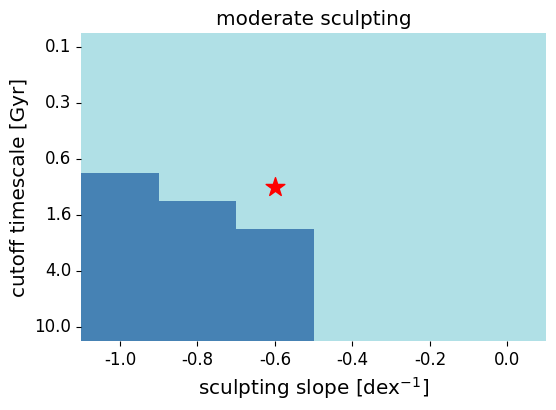}&
    \includegraphics[width=0.28\textwidth]{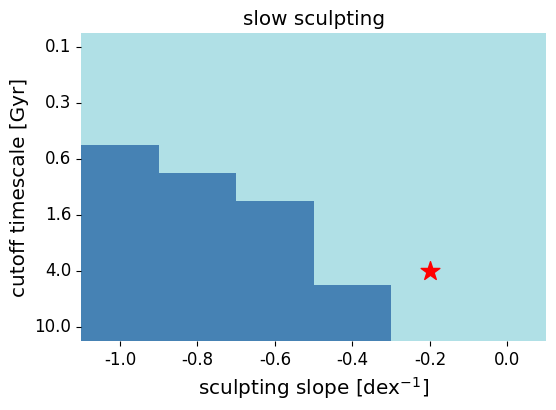}&
    \includegraphics[width=0.28\textwidth]{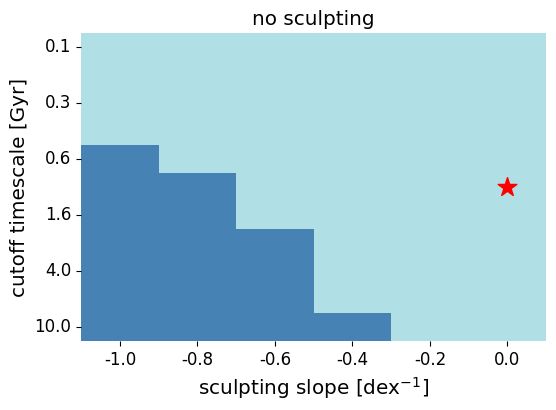}
  \end{tabular}}
  \rotatebox{270}{\begin{minipage}[c][\textwidth][c]{\wd0}
    \usebox0
    \caption{We show the same series of four injection-recovery tests depicted in Figure \ref{fig:injection-recovery-fig}, but with loguniformly distributed mean ages and uniformly distributed 10\% age errors, instead of those derived from isochrones and provided in the \citet{berger_gaia-kepler_2020} \textit{Gaia}-\textit{Kepler} cross-match. We find that slightly less of the free parameter space is ruled out by the pipeline, but most of the prominent features noted in the previous injection-recovery tests (eg. disfavored low \init{} if the ground truth is not low \init{}) persist. One feature that does change is that the lower right corner of shallow sculpting laws with high intact fractions at 0.1 Gyr can no longer be ruled out among moderate and slower sculpting laws.}
    \label{fig:injection-recovery-loguniform}
  \end{minipage}}
\end{figure*}

We demonstrated in Section \ref{sec:strong_age_constraint} that stellar age precision is not the limiting factor in our ability to recover $\Theta$. We investigate one final possibility: that a set of stars with log uniform age sampling -- particularly one that explicitly addresses our dearth of stars between 0.1 and 1 Gyr old -- could enable us to make stronger claims about sculpting timescales. We therefore conduct another injection-recovery experiment using log uniform ages for just the G dwarfs in the sample (since a 10 Gyr-old F star or a 0.1 Gyr-old K star is highly unlikely) by dividing the culled sample into ten equally-sized bins from 0.1 to 10 Gyr in log space. As in the previous injection-recovery test, 10\% age errors are painted onto the sample, once ages are re-assigned. We show our findings in Figure \ref{fig:injection-recovery-loguniform}. We find that the forward modeling pipeline recovers approximately the same present-day intact fraction. Furthermore, general trends we have identified in the previous injection-recovery tests appear to remain true: $\Theta$ that underpredict the number of observed multis correspond to zero-to-low \init{}, ``fast" sculpting in which the intact fraction decays rapidly to zero, and long-timescale sculpting in which no intact systems remain. One difference between the results of this injection-recovery test and the previous ones is that the set of models corresponding to a high intact fraction at 0.1 Gyr and a shallow sculpting law is no longer ruled out among moderate and slower sculpting ground truths.

\section{Conclusions}
\label{sec:Conclusions}

We have investigated here the sensitivity of the \textit{Kepler} transit multiplicity as a tracer of Gyr-timescale dynamical sculpting. Our study is motivated by suggestive recent investigations of planet occurrence in a galactic context. Specifically, a finding by \citet{chen_planets_2021} shows an offset between the ages of hosts stars to singly-transiting, versus multi-transiting systems. Multi-transiting systems appear to be 1.5 Gyr younger on average; one plausible scenario is that initially flat and compact planetary systems evolve over Gyr timescales. If this hypothesized dynamical sculpting is sufficient to remove planets from the transit geometry, the effect would be hypothetically observable via this change to the transit multiplicity distribution as a function of stellar age. 

To test this hypothesis, we posit a suite of toy sculpting models, in which some planetary systems are ``born" dynamically cold and ``intact" (as so-called ``systems with tightly-packed inner planets", or STIPs), and others dynamically hot, at a starting point of 10$^{8}$ years. We then model the likelihood of an ``intact" system becoming ``disrupted" over time. We require only that such a process take (1) the form of a power law and (2) increase the dynamical temperature of planetary systems over time. We parameterized each sculpting law with four variables: the initial ``intact" fraction of systems at 0.1 Gyr, the rate at which ``intact" systems become ``disrupted", the time at which this hypothesized sculpting ceases, and the fraction of stars that host a planetary system of some kind. With these sets of parameters $\Theta$, we then synthesize planetary systems to population the \textit{Kepler} host stars, enabled by stellar age constraints for FGK dwarfs from \textit{Gaia}. We then investigate the likelihood of observing the \textit{Kepler} transit multiplicity given $\Theta$. Our findings include the following:

\begin{itemize}
  \item From an injection-and-recovery analysis, we demonstrate the ability to recover ground truth sculpting laws $\Theta$ across different regimes. However, these $\Theta$ are not unique: sculpting laws that correctly reproduce the fraction of STIPs are statistically indistinguishable from one another. In this sense, we rule out parts of $\Theta$ parameter space that produce too \textit{many} multi-transit systems, and those that produce too \textit{few}.
  \item Upon comparison of the predictions of our sculpting models to the real \textit{Kepler} data, we find that the data support $\Theta$ that result in STIP-type systems orbiting between 4 and 13\% of FGK dwarfs. Our likeliest fraction of stars hosting any type of planetary system is 0.4 to 0.5, consistent with other estimates from the literature. 
  \item The $\Theta$ that are inconsistent with the \textit{Kepler} transit yield include: dynamical processes that proceed too quickly (steep change in excitation likelihood/dex in age) and/or over too long of a timescale (i.e. for Gyrs), as the predicted number of intact systems is too small to be consistent with the observing multi-planet systems in the data. Similarly, intact fractions at 0.1 Gyr must always be $\geq 0.1$, or else the model underpredicts the number of multis. Conversely, sculpting laws that predict high rates of STIPs at 10$^{8}$ years (starting ``intact" fractions $\geq 0.3$, which then decay slowly or not at all) are ruled out because they furnish too many predicted multi-transit systems. 
  \item There exists no sculpting law $\Theta$, consistent with the \textit{Kepler} transit multiplicity yield today, that produces an observable offset in age between systems with 0 or 1 transiting planets and those with multiple transiting planets. We demonstrate that improving the precision for stellar age measurements can furnish at least one model yielding a transit multiplicity close to that observed by \textit{Kepler}, while also producing a noticeable difference in age between systems with 0 or 1 planets and those with 3 or more planets. For most models that closely match the \textit{Kepler} transit multiplicity in this idealized scenario, however, intra-system contamination still curbs the production of a measurable offset between the samples (that is, the fact that a large fraction of systems hosting ``singles" or no transiting planets are, in fact, STIPS, means that the samples overlap substantially). 

  
\end{itemize}


We consider our last finding, that (within our set of assumptions and the current precision of stellar ages in our sample) Gyr-timescale sculpting cannot reproduce an offset in age between ``singles" and ``multis".  \citet{chen_planets_2021} folded in data from the Large Sky Area Multi-Object Fiber Spectroscopic Telescope (LAMOST), along with \textit{Kepler} and \textit{Gaia}, to find a trend of increasing transit multiplicity with lower stellar kinematic speed and younger age. We conclude that other mechanisms must likely be operative to produce this offset. Stellar age, rather than serving as a clock for dynamical sculpting, may be a proxy instead for the dynamical history of the star itself. In this sense, stellar kinematic history may be more deterministic \citep{Winter20, dai_planet_2021, chen_planets_2021, Longmore21, bashi_exoplanets_2022, zink_scaling_2023} if planet occurrence can be linked to galactic environment directly. Both \cite{Cai17} and \cite{Ndugu22} have proposed mechanisms by which stellar encounters affect planetary systems. Alternatively, any age offset with transit multiplicity may reflect underlying relationships with \textit{galactic} metallicity \citep{Nielsen23}, though \cite{zink_scaling_2023} found that the predicted shift toward higher metallicity with age in the Milky Way cannot entirely explain a change in planet occurrence with galactic height.

Among the dynamical sculpting models that reproduce the \textit{Kepler} transit multiplicity are models in which \textit{no} sculpting occurs at Gyr timescales: that is, the imprints of any dynamical sculpting on the \textit{Kepler} transit multiplicity are fixed at 100 Myrs, from some combination of planet formation, the galactic environment, and non-secular, intra-system dynamical processes. Whether or not, to what extent, at which timescales, and by which mechanisms these systems change post-formation will have very different implications for habitability on worlds around stars similar to our own. Such a scenario would not \textit{a priori} explain an offset in ages between singles and multis; on the other hand, \citet{dai_planet_2021} did not recover the same offset in age between hosts to single- and multi-transit systems. Rather, they found the opposite relationship, noting that the multi-to-single ratios of low-velocity stars were \textit{lower} -- and their eccentricities higher -- than those of higher-velocity stars, reasoning that higher-velocity stars may undergo fewer disruptive events from their Galactic environment. As our sample of planets grows to include  kinematically distinct samples of host stars (from e.g. the thin disk and thick disk), we will better understand planet formation within a galactic context (see e.g. \citealt{zink_scaling_2023}).

\section{Acknowledgments}
\label{sec:acknowledgments}
This research has made use of the NASA Exoplanet Archive, which is operated by the California Institute of Technology, under contract with the National Aeronautics and Space Administration under the Exoplanet Exploration Program. I will include a full person and land acknowledgment in the final draft. This material is based upon work supported in part by the National Science Foundation GRFP under Grant No. 1842473. We wish to thank Jamie Tayar, Matthias He, Jason Dittmann, Quadry Chance, Sheila Sagear, Natalia Guerrero, and especially Sarah Millholland for their helpful comments and suggestions. We acknowledge that for thousands of years the area now comprising the state of Florida has been, and continues to be, home to many Native Nations. We further recognize that the main campus of the University of Florida is located on the ancestral territory of the Potano and of the Seminole peoples. The Potano, of Timucua affiliation, lived here in the Alachua region from before European arrival until the destruction of their towns in the early 1700s. The Seminole, also known as the Alachua Seminole, established towns here shortly after but were forced from the land as a result of a series of wars with the United States known as the Seminole Wars. We, the authors, acknowledge our obligation to honor the past, present, and future Native residents and cultures of Florida.

\facility{Exoplanet Archive, Kepler, Gaia}


\bibliography{Sculpting,sculpting_additional}
\bibliographystyle{aasjournal}

\end{document}